\documentclass[sigconf]{acmart}

\setcopyright{none} 

\usepackage{amsmath,amsfonts}
\usepackage{graphicx}
\usepackage{textcomp}
\usepackage{xcolor}
\usepackage{algorithmicx}
\usepackage{algpseudocode}
\usepackage{xspace}
\usepackage{tabularx}
\usepackage{longtable}
\usepackage{multirow}
\usepackage{subcaption}
\usepackage{algorithm}
\usepackage{algpseudocode}
\usepackage{todonotes}
\usepackage{graphicx}
\usepackage{xcolor}
\usepackage{wrapfig}

\usepackage{enumitem}
\usepackage{tikz}
\usepackage{longtable}
\usepackage[normalem]{ulem}

\newcommand{\ourname}{\textsc{DAWN}\xspace}
\newcommand{\adv}{$\mathcal{A}$\xspace}
\newcommand{\victim}{$\mathcal{V}$\xspace}
\newcommand{\judge}{$\mathcal{J}$\xspace}
\newcommand{\client}{API client\xspace}

\newcommand{\oldtext}[1]{\textcolor{black}{#1}}

\newcommand{\accallfail}[1]{\underline{\textcolor{violet}{#1}}}
\newcommand{\accwmfail}[1]{\dashuline{\textcolor{red}{#1}}}
\newcommand{\techreport}[1]{}

\newcommand*\circled[1]{\tikz[baseline=(char.base)]{
            \node[shape=circle,fill,inner sep=1pt] (char) {\textcolor{white}{#1}};}}

\begin{document}

\title{\ourname: Dynamic Adversarial Watermarking of Neural Networks}

\author{Sebastian Szyller}
\affiliation{%
  \institution{Aalto University}
  \country{Finland}}
\email{contact@sebszyller.com}

\author{Buse Gul Atli}
\affiliation{%
  \institution{Aalto University}
  \country{Finland}}
\email{buse.atlitekgul@aalto.fi}

\author{Samuel Marchal}
\affiliation{%
  \institution{Aalto University \& F-Secure Corporation}
  \country{Finland}}
\email{samuel.marchal@aalto.fi}

\author{N. Asokan}
\affiliation{%
  \institution{University of Waterloo \& Aalto University}
  \country{Canada}}
\email{asokan@acm.org}

\begin{abstract}
    Training machine learning (ML) models is expensive in terms of computational power, amounts of labeled data and human expertise.
    Thus, ML models constitute intellectual property (IP) and business value for their owners. Embedding digital watermarks during model training allows a model owner to later identify their models in case of theft or misuse. However, model functionality can also be stolen via \emph{model extraction}, where an adversary trains a \emph{surrogate model} using results returned from a prediction API of the original model. Recent work has shown that model extraction is a realistic threat. Existing watermarking schemes are ineffective against IP theft via model extraction since it is the adversary who trains the surrogate model.
    In this paper, we introduce \ourname (Dynamic Adversarial Watermarking of Neural Networks), the first approach to use watermarking to deter model extraction IP theft.
    Unlike prior watermarking schemes, \ourname does not impose changes to the training process but
    it operates at the prediction API of the protected model, by dynamically changing the responses for a small subset of queries (e.g., \textless 0.5\%) from API clients.
    This set is a watermark that will be embedded in case a client uses its queries to train a surrogate model.
    We show that
    \ourname is resilient against two state-of-the-art model extraction attacks, effectively watermarking all extracted surrogate models, allowing model owners to reliably demonstrate ownership (with confidence \textgreater $1- 2^{-64}$), incurring negligible loss of prediction accuracy (0.03-0.5\%).
\end{abstract}


\ccsdesc[500]{Security and privacy~Systems security}
\ccsdesc[500]{Computing methodologies~Machine learning}

\keywords{Deep Neural Network, Watermarking, Model stealing, IP protection}

\maketitle

\section{Introduction}
\label{Introduction}

Recent progress in machine learning (ML) has led to a dramatic surge in the use of ML models for a wide variety of applications. Major enterprises like Google, Apple, and Facebook have already deployed ML models in their products~\cite{techworld:2018}. ML-related businesses are expected to generate trillions of dollars in revenue in the near future~\cite{forbes:2019}. The process of collecting training data and training ML models is the basis of the business advantage of model owners.
Hence, protecting the intellectual property (IP) embodied in ML models is necessary. 

One approach for IP protection of ML models is \emph{watermarking}. Recent work~\cite{merrer2017adversarial,adi2018turning,zhang2018protecting} has shown how \emph{digital watermarks} can be embedded into deep neural network models (DNNs) during training. Watermarks consist of a set of inputs, the \emph{trigger set}, with incorrectly assigned labels. A legitimate model owner can use the trigger set, along with a large training set with correct labels, to train a watermarked model and distribute it to his customers. If he later encounters a model he suspects to be a copy of his own, he can demonstrate ownership by using the trigger set as inputs to the suspected model. These watermarking schemes allow legitimate model owners to detect
theft or misuse of their models.

Instead of distributing ML models to customers, an increasingly popular alternative business paradigm is to allow customers to use models via \emph{prediction APIs}. But one can mount a \emph{model extraction}~\cite{tramer2016stealing} attack via such APIs by sending a sequence of API queries with different inputs and using the resulting predictions to train a \textit{surrogate model} with similar functionality as the queried model. Model extraction attacks are effective even against complex DNN models~\cite{juuti2019prada,orekondy2018knockoff}, and are difficult to prevent~\cite{juuti2019prada}.
Existing watermarking techniques, which rely on model owners to embed watermarks during training, are  ineffective against model extraction since it is the adversary who trains the surrogate model.

In this paper we introduce \ourname (Dynamic Adversarial Watermarking of Neural Networks), a new watermarking approach intended to deter IP theft via model extraction. \ourname is designed to be deployed within the prediction API of a model.
It dynamically watermarks a tiny fraction of queries from a client by changing the prediction responses for them. The watermarked queries serve as the trigger set if an adversarial client trains a surrogate model using the responses to its queries. The model owner can use the trigger set to demonstrate IP ownership of the extracted surrogate model as in prior DNN watermarking solutions~\cite{adi2018turning,zhang2018protecting}.
\ourname differs from them in that it is the adversary (model thief), rather than the defender (original owner) who trains the watermarked model. This raises two new challenges: (1) defenders must choose trigger sets from among queries sent by clients and cannot choose optimal trigger sets from the whole input space; (2) adversaries can select the training data or manipulate the training process to resist the embedding of watermarks.
\ourname addresses both these challenges.

\ourname watermarks are client-specific: \ourname not only infers whether a given model is a surrogate but, in case of model extraction, also identifies the client whose queries were used to train the surrogate. \ourname is parametrized so that changed predictions needed for watermarking are sufficiently rare as to not degrade the utility of the original model for legitimate API clients.

We make the following contributions:
\begin{itemize}
\item present \ourname, the first approach for dynamic, selective watermarking for DNN models at their prediction APIs for deterring IP theft via model extraction (Sect.~\ref{sec:approach}),
  	\item empirically assess it (Sect.~\ref{sec:exp-setup}) using several DNN models and datasets showing that \ourname is robust to adversarial manipulations and resilient to evasion (Sect.~\ref{sec:eval-perfect} and~\ref{sec:watermark-removal}), and
	\item show that \ourname is resistant to two state-of-the-art extraction attacks, reliably demonstrating ownership (with confidence \textgreater $1- 2^{-64}$) with negligible impact on model utility (0.03-0.5\% decrease in accuracy) (Sect.~\ref{sec:eval-stealing}).
\end{itemize}

Code to reproduce our experiments is available on GitHub\footnote{
	\textcolor{blue}{
		\href{https://github.com/ssg-research/dawn-dynamic-adversarial-watermarking-of-neural-networks}{github.com/ssg-research/dawn-dynamic-adversarial-watermarking-of-neural-networks}
	}
}.

\section{Background}
\label{related-work}

\techreport{\subsection{Deep Neural Network}

A DNN is a function $F: \mathbb{R}^n \rightarrow \mathbb{R}^m$, where $n$ is the number of input features and $m \ge 2$ is the number of output classes in a classification task.
$F(x)$ is a vector of length $m$ containing probabilities $p_j$ that $x$ belongs to each class $c_j \in C$ for $j \in \lbrace 1, m\rbrace$.
The predicted class, $\hat{F}(x)$, is obtained by applying the \textit{argmax} function: $\hat{F}(x) = argmax(F(x)) = c$.
$F$ is trained such that $\hat{F}(x)$ approximates a perfect oracle function $O_f: \mathbb{R}^n \rightarrow C$ which gives the true class $c$ for any sample $x \in \mathbb{R}^n$. If $F$ is successfully trained, $\hat{F} \sim O_f$, and the accuracy of $F$ is close to 1: $Acc(F) = 1 - \epsilon$ where $\epsilon$ is the irreducible error.
}

\subsection{Model Extraction Attacks}
\label{sec:bg_extraction}

In model extraction~\cite{tramer2016stealing, juuti2019prada, orekondy2018knockoff, papernot2017practical, correia2018copycat, pal2020activethief}, an adversary \adv wants to ``steal'' a DNN model $F_\mathcal{V}$ of a victim \victim by making a series of prediction requests $U$ to $F_\mathcal{V}$ and obtaining predictions $F_\mathcal{V}(U)$.
$U$ and $F_\mathcal{V}(U)$ are used by \adv to train a surrogate model $F_\mathcal{A}$. \adv's goal is to have $Acc(F_\mathcal{A})$ as close as possible to $Acc(F_\mathcal{V})$.
All model extraction attacks~\cite{tramer2016stealing, juuti2019prada, orekondy2018knockoff, papernot2017practical, correia2018copycat} operate in a \emph{black-box} setting: \adv has access to a prediction API, \adv uses the set $<U, F_\mathcal{V}(U)>$ to iteratively refine the accuracy of $F_\mathcal{A}$.
Depending on the adversary model, \adv's capabilities can be divided into three categories: \textit{model knowledge}, \textit{data access}, and \textit{querying strategy}.

\textbf{Model knowledge.} \adv does not know the exact architecture of $F_\mathcal{V}$ or the hyperparameters or the training process.
However, given the purpose of the API (e.g., image recognition) and expected complexity of the task, \adv may attempt to guess the architecture of the model~\cite{papernot2017practical, juuti2019prada}.
On the other hand, if $F_\mathcal{V}$ is complex, \adv can use a publicly available, high capacity model pre-trained with a very large
benchmark datasets~\cite{orekondy2018knockoff}.
While the above methods focus on DNNs, there are alternatives targeting simpler models: logistic regression, decision trees, shallow neural networks~\cite{tramer2016stealing}.

\textbf{Data access.} \adv's main limitation is the lack of access to \emph{natural data} that comes from the \emph{same distribution} as the data used to train $F_\mathcal{V}$.
\adv may use data that comes from the same domain as \victim's training data but from a different distribution~\cite{correia2018copycat}.
If \adv does not exactly know the distribution or the domain, it may use widely available natural data~\cite{orekondy2018knockoff, pal2020activethief} to mount the attack. Alternatively, it may use only \emph{synthetic samples}~\cite{tramer2016stealing} or a mix of a small number of natural samples augmented by synthetic samples~\cite{papernot2017practical, juuti2019prada}.

\textbf{Querying strategy.} All model stealing attacks~\cite{tramer2016stealing, juuti2019prada, orekondy2018knockoff, papernot2017practical,  correia2018copycat,pal2020activethief} consist of alternating phases of \adv querying $F_\mathcal{V}$, followed by training the surrogate model $F_\mathcal{A}$ using the obtained predictions.
\adv queries $F_\mathcal{V}$ with all its data and then trains the surrogate model~\cite{correia2018copycat,orekondy2018knockoff}.
Alternatively, if \adv relies primarily on synthetic data~\cite{papernot2017practical, juuti2019prada}, it deliberately crafts inputs that would help it train $F_\mathcal{A}$.

\subsection{Watermarking DNN models}
\label{sec:bg_wm}

Digital watermarking is a technique used to covertly embed a marker, \emph{the watermark}, in an object (image, audio, etc.) which can be used to demonstrate ownership of the object.
Watermarking of DNN models leverages the massive overcapacity of DNNs and their ability to fit data with arbitrary labels~\cite{zhang2016understanding}.
DNNs have a large number of parameters, many of which have little significance for their primary classification task. These parameters can be used to carry additional information beyond what is required for its primary classification task.
This property is exploited by backdooring attacks, which consist in training a DNN model that deliberately outputs incorrect predictions for some selected inputs~\cite{chen2017targeted,gu2017badnets}.

Watermarking of DNN models is currently based on backdooring attacks~\cite{merrer2017adversarial,adi2018turning,zhang2018protecting,DarvishRouhani:2019:DEW:3297858.3304051}.
We want to train a DNN model $F: \mathbb{R}^n \rightarrow \mathbb{R}^m$ for which the primary task is that $\hat{F}(x) = argmax(F(x)) = c$ approximates an oracle $O_f: \mathbb{R}^n \rightarrow C$.
Embedding a watermark in $F$ consists of enabling $F$ with a secondary classification task: for a subset of samples $x \in T \subset \mathbb{R}^n$, we want $\hat{F}$ to output incorrect prediction classes as defined by a function $B: T \rightarrow \mathbb{R}^m$ such that $\hat{B}(x) \neq O_f(x)$.
We call $B(x)$ a \textit{backdoor function} and $T$ a \textit{trigger set}: $T$ triggers the backdoor.
$F$ is trained using the trigger set $T$ mislabeled using $\hat{B}(x)$ in addition to a larger set of samples $x \in \mathbb{R}^n \setminus T$ accurately labeled using $O_f(x)$.
$F$ is a watermarked DNN model which is expected to approximate the backdoor function $B(x)$ for $x \in T$ and the oracle $O_f$ for $x \in \mathbb{R}^n \setminus T$.

The trigger set $T$ and the outputs of the backdoor function for its elements $\hat{B}(T)$ compose the watermark: $(T,\hat{B}(T))$.
Let $F'$ be a DNN model that copies $F$.
The watermark can be used to demonstrate ownership of $F'$.
It only requires $F'$ to expose a prediction API which can be used to query all samples in the trigger set $x \in T$.
A sufficient number of predictions $\hat{F}'(x)$ such that $\hat{F}'(x) = \hat{B}(x)$ demonstrates that $F'$ is a copy of the watermarked model $F$.

\section{Problem Statement}
\label{problem-statement}


\subsection{Adversary Model}
\label{sec:adv_model}

The adversary \adv mounts a model extraction attack against a victim model $F_\mathcal{V}$ using queries to its prediction API.
\adv's goal is \textit{model functionality stealing}~\cite{orekondy2018knockoff}: train a surrogate model $F_\mathcal{A}$ that performs well on a classification task for which $F_\mathcal{V}$ was designed. If $\hat{F}_\mathcal{V} \sim O_f$ then \adv's goal is that $\hat{F}_\mathcal{A} \sim O_f$, which can be considered successful if $Acc(F_\mathcal{A}) \sim Acc(F_\mathcal{V})$.
A secondary goal is to minimize the number of queries to $F_\mathcal{V}$ necessary for \adv to train $F_\mathcal{A}$.

\adv has full control over the samples $D_\mathcal{A}$ it chooses to query $F_\mathcal{V}$ with. These can be natural~\cite{orekondy2018knockoff} or synthetic~\cite{juuti2019prada,papernot2017practical,tramer2016stealing}.
\adv obtains a prediction for each query in the form of probability vectors $F_\mathcal{V}(x)$ or single classes $\hat{F}_\mathcal{V}(x), \forall x \in D_\mathcal{A}$.
\adv uses queried samples and their predictions to train $F_\mathcal{A}$, a DNN.
It chooses the DNN model architecture, training hyperparameters and training process.
\oldtext{Requiring $F_\mathcal{A}$ to be a DNN is justified by the observations in prior work on model extraction attacks~\cite{juuti2019prada,orekondy2018knockoff,papernot2017practical} that $F_\mathcal{A}$ needs to have equal or larger capacity than $F_\mathcal{V}$ in order for model extraction to be successful. DNNs have the greatest capacity among ML models~\cite{zhang2016understanding}.}

\subsection{Assumptions}
\label{sec:assumption}

We assume that for a given input $x \in D_\mathcal{A}$, \adv has no a priori expectation regarding the prediction $F_\mathcal{V}(x)$.
\adv treats $y = F_\mathcal{V}(x)$ as the ground truth label for $x \in D_\mathcal{A}$.
\adv expects that multiple queries of the same input $x$ must return the same prediction $y$.

Our focus is on \adv who makes $F_\mathcal{A}$ available via a prediction API since it has the greatest impact on \victim's business advantage.
We do not consider an \adv who keeps $F_\mathcal{A}$ for private use.
This is similar to media watermarking schemes where access to allegedly stolen media is a pre-requisite for ownership demonstration~\cite{petitcolas1999information}.

\subsection{\ourname Goals and Overview}
\label{sec:dawn_overview}

On one hand, model extraction attacks against DNNs have been proven difficult to defend against~\cite{juuti2019prada}.
On the other hand, existing watermarking techniques~\cite{merrer2017adversarial,adi2018turning,DarvishRouhani:2019:DEW:3297858.3304051,chen2018deepmarks,li2019persistentwatermarks} are vulnerable to model extraction attacks~\cite{zhang2018protecting}.
To address these limitations, we design a solution to identify and prove the ownership of DNN models stolen through a prediction API.

\begin{figure}[th]
                \centering
                \includegraphics[width=0.95\columnwidth]{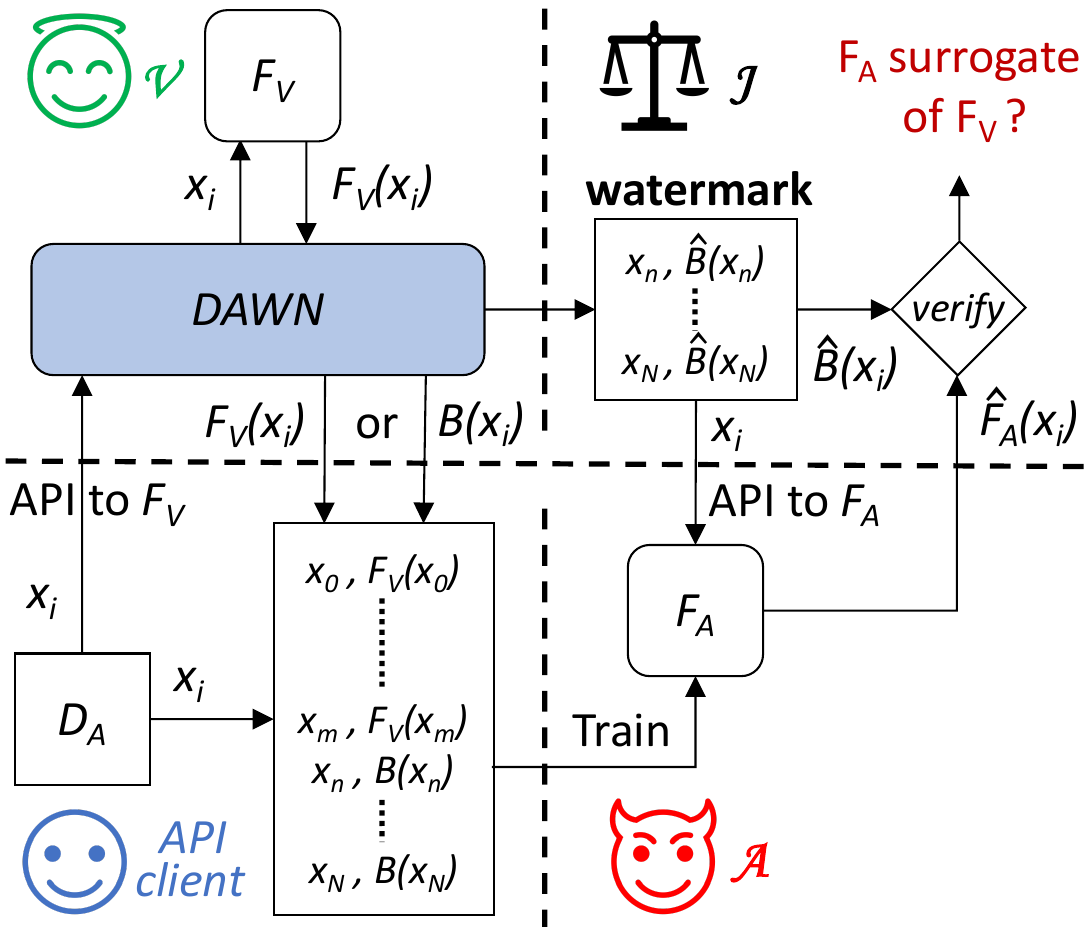}
                \caption{\ourname system overview with four parties: a victim \victim owning a model $F_\mathcal{V}$, \client{s} querying the model prediction API, an adversary \adv training a surrogate model $F_\mathcal{A}$ and a judge \judge verifying the surrogacy of $F_\mathcal{A}$.}
                \label{fig:dawn-oveview}
\end{figure}

Our solution, \ourname (Dynamic Adversarial Watermarking of Neural Networks), is an additional component added in front of a model prediction API (Fig.~\ref{fig:dawn-oveview}).
\ourname dynamically embeds a watermark in responses to queries made by a \textit{\client}.
This watermark is composed of inputs $x_i \in T$ for which we return incorrect predictions $B(x_i) \neq F_\mathcal{V}(x_i)$.
\adv uses all the responses including these mislabeled samples $(x_i, B(x_i))$ to train $F_\mathcal{A}$.
$F_\mathcal{A}$ will remember those samples as a backdoor~\cite{chen2017targeted} that represents the watermark (as in traditional DNN watermarking techniques).
If $F_\mathcal{A}$ exposes a public prediction API, a judge \judge can run a verification process (\textit{verify}), which confirms $F_\mathcal{A}$ is a surrogate of $F_\mathcal{V}$. \textit{Verify} checks that for sufficient number of inputs $x_i \in T$, we have $\hat{F}_\mathcal{A}(x_i) = \hat{B}(x_i) \neq \hat{F}_\mathcal{V}(x_i)$.
\ourname embeds a watermark into a subset of queries it receives so that any $F_\mathcal{A}$ trained using these responses will retain the watermark.

\subsection{System requirements}
\label{sec:system-requirements}

We define the following requirements for the watermark that \ourname embeds in $F_\mathcal{A}$ during an extraction attack. \textbf{W1-W3} were introduced in~\cite{adi2018turning} while \textbf{W4} is a new requirement specific to \ourname.

\begin{enumerate}[label=\textbf{W\arabic*}]
  \item \label{req:unremovability} \textbf{Unremovability}: \adv is unable to remove the watermark from $F_\mathcal{A}$ without significantly decreasing its accuracy, rendering it ``unusable''. If $F_\mathcal{V}$ is free of the watermark, then $Acc(F_\mathcal{A}) \ll Acc(F_\mathcal{V})$.
    \item \label{req:non-trivial} \textbf{Reliability}: If \textit{verify} outputs ``true'' for a watermark $(T,\hat{B}_{\mathcal{V}}(T))$ on a model $F'$, then $F'$ is a surrogate of  $F_\mathcal{V}$, with high confidence. On the other hand, if $F'$ is not a surrogate, \adv cannot generate a watermark $(T,\hat{B}(T))$ such that \textit{verify} outputs ``true''  (non-trivial ownership).
    \item \label{req:non-piracy} \textbf{Non-ownership piracy}: \adv cannot produce a watermark for a model that was already watermarked by \victim, such that it can cast \victim's ownership into doubt.
    \item \label{req:linkability} \textbf{Linkability}:
If \textit{verify} outputs ``true'' for a model $F_\mathcal{A}$, the watermark used for verification $(T,\hat{B}(T))$ can be linked to a specific \client whose queries were used to train $F_\mathcal{A}$.
\end{enumerate}

We identify additional requirements \ref{req:qos}-\ref{req:sybils}:
\begin{enumerate}[label=\textbf{X\arabic*}]
  \item \label{req:qos} \textbf{Utility}: Incorrect predictions returned by \ourname do not significantly degrade the prediction service provided to legitimate \client{s}: $Acc(\ourname + F_\mathcal{V}) \sim Acc(F_\mathcal{V})$.
   \item \label{req:indifferentiability} \textbf{Indistinguishability}: \adv cannot distinguish incorrect predictions $B(x)$ from correct victim model predictions $F_\mathcal{V}(x)$.
    \item \label{req:sybils} \textbf{Collusion resistance}: Watermark unremovability (\ref{req:unremovability}), linkability (\ref{req:linkability}) and indistinguishability (\ref{req:indifferentiability}) must remain valid even if the extraction attack is distributed among several \client{s}.
\end{enumerate}

\techreport{In contrast to traditional DNN watermarking, \ourname does not aim at maximizing the accuracy of the watermarked model $Acc(F_\mathcal{A})$ on a primary classification task.}

\subsection{Relation to other attacks}
\label{sec:other-attack}

\ourname is different from prior work where the goal is to (a) degrade model performance (decrease test accuracy $Acc_{test}$ -- typical of poisoning attacks~\cite{biggio2012poisoning,munoz2017towards}), (b) trigger targeted misclassifications (classify a trigger set with high accuracy $Acc_{bd}$ -- typical to backdooring~\cite{liu2018trojaning}) or (c) embed a watermark while preserving high model performance (reach high $Acc_{bd}$ and $Acc_{test}$ -- typical of DNN watermarking~\cite{merrer2017adversarial,adi2018turning}).
In contrast to backdooring, \ourname cannot inject arbitrary samples in $D_\mathcal{A}$ but it  can modify the label of $D_\mathcal{A}$ samples to any incorrect prediction $c \neq \hat{F}_\mathcal{V}(x)$. 
In contrast to traditional DNN watermarking, \victim neither controls the training of $F_\mathcal{A}$ nor can it choose the trigger set from the whole input space $\mathbb{R}^n$: \victim is limited to the set of samples $D_\mathcal{A}$ submitted by \adv. Table~\ref{tab:comparison} summarizes these differences.

\begin{table}[htb]
\begin{center}
	\caption{Adversarial watermarking (\ourname) capabilities and goals compared to (a) poisoning attacks, (b) backdoor attacks and (c) prior DNN watermarking.}
	\label{tab:comparison}
	\begin{tabular}{lccccc} \hline
							& 	\multicolumn{3}{c}{\textbf{Capabilities}} & \multicolumn{2}{c}{\textbf{Goal}}				 \\
					& 	Modify   &  Inject  & Control 	&  & \\
						& 	labels  &   in  $D_\mathcal{A}$ &  training &  $Acc_{test}$ & $Acc_{bd}$  \\ \hline
		Poisoning			& Yes		& Yes 	& No					& Low & --  \\
		Backdoor			& Yes		& Yes				& Yes / No	& -- & High  \\
		Watermarking	& Yes		& Yes				& Yes				& High  & High   \\
		\textbf{\ourname}			& Yes		& \textbf{No}					& \textbf{No}					& \textbf{--} & High\\  \hline
	\end{tabular}
\end{center}
\end{table}

\section{Dynamic Adversarial Watermarks}
\label{sec:approach}

We first present the method for generating and embedding an adversarial watermark. Then we describe the process for proving ownership of a model using the watermark.

\subsection{Watermark generation}
\label{sec:wm-generation}

We define \textit{watermarking an input $x$} as returning an incorrect prediction $B_\mathcal{V}(x)$ instead of the correct prediction $F_\mathcal{V}(x)$.
The collection of all watermarked inputs composes the trigger set $T_\mathcal{A}$ that will be a backdoor to any $F_\mathcal{A}$ trained using responses from $F_\mathcal{V}$ including $T_\mathcal{A}$.
Consequently, inputs $x \in T_\mathcal{A}$ and their corresponding prediction classes $\hat{B}_\mathcal{V}(x)$ compose the watermark to the surrogate model $(T_\mathcal{A},\hat{B}_\mathcal{V}(T_\mathcal{A}))$.
We define two functions:
\begin{itemize}
    \item $W_\mathcal{V}(x)$: should the response to $x$ be watermarked?
    \item $B_\mathcal{V}(x)$: what is the (backdoored watermark) response? 
\end{itemize}

\adv must not be able to predict $W_\mathcal{V}(x)$ 
or distinguish between $B_\mathcal{V}(x)$ and $F_\mathcal{V}(x)$.
The same query, regardless of the \client, must always get the same output.
Both functions must be \textit{deterministic} random functions specific to $F_\mathcal{V}$ to fulfill these properties.

We use the result of a keyed cryptographic hash function as a source for randomness.
We compute $\text{HMAC}(K_w,x)$ using SHA-256, where $K_w$ is a model-specific secret key generated by \ourname and $x$ is an input to $F_\mathcal{V}$.
If $x$ is a matrix of dimension $d > 1$, it is flattened to a 1-dimensional vector.
The result of the hash is split in two parts $\text{HMAC}(K_w,x)[0,127]$ and $\text{HMAC}(K_w,x)[128,255]$, respectively used in $W_\mathcal{V}$ and $B_\mathcal{V}$.
These numbers are independent and provide a sufficient source for randomness for each function.

\subsubsection{Watermarking decision}

$W_\mathcal{V}(x)$ is a boolean function. 
We define $r_w$ as the fraction of inputs to be watermarked out of $N$ inputs submitted by an \client.
$r_w$ will define the size of the trigger set $\vert T_\mathcal{A} \vert = \lfloor r_w \times N \rfloor $.
Then:

\begin{equation}
    W_\mathcal{V}(x) = \begin{cases}
    1, & \text{if  HMAC$(K_w,x)[0,127] < r_w \times 2^{128}$}.\\
    0, & \text{otherwise}.
  \end{cases}
\end{equation}

The expectation that $W_\mathcal{V}$ returns $1$  and thus to watermark a sample is uniformly equal to $r_w$.
It is worth noting that \ourname does not differentiate adversaries from benign \client{s}. Consequently, any \client obtains a rate $r_w$ of incorrect predictions.
$r_w$ must be defined to meet a trade-off.
A large $r_w$ increases the reliability of ownership demonstration and prevents trivial ownership demonstration \ref{req:non-trivial} as later discussed in Sect.~\ref{sec:own-proof}. A small $r_w$ maximizes utility \ref{req:qos} by minimizing the number of incorrect predictions returned to benign \client{s}.

\subsubsection{Backdoor function}

We implement the backdoor function $B_\mathcal{V}(x)$ as a function of $F_\mathcal{V}(x)$. Our motivations are two-fold. First, this allows for deploying \ourname to protect any model $F_\mathcal{V}$ without the need for redefining $B_\mathcal{V}$. Second, it makes $B_\mathcal{V}(x)$ consistent with correct predictions $F_\mathcal{V}(x)$.
We define $B_\mathcal{V}(x) = \pi(K_\pi,F_\mathcal{V}(x))$ where $\pi: \mathbb{R}^m \rightarrow \mathbb{R}^m$ is a keyed pseudo-random permutation function with secret key $K_\pi$.
Even if an adversary uncovers values $B_\mathcal{V}(x)$ for a large number of inputs, it will not be able to infer the function $B_\mathcal{V}$. This prevents an adversary from recovering $F_\mathcal{V}(x)$ from $B_\mathcal{V}(x)$ in case it knows if an input is backdoored.

$\pi(K_\pi,F_\mathcal{V}(x))$ does not need to permute all $m$ positions of $F_\mathcal{V}(x)$ but only those with highest probabilities for the purpose of backdooring. A large number of classes typically have a 0 probability value when $m$ is large.
Considering that the number of positions to permute is small, we use the Fisher-Yates shuffle algorithm~\cite{fisher1949statistical} to implement $\pi$.
We use $K_\pi = \text{HMAC}(K_w,x)\left[128,255\right]$ as the key that determines the permutations performed during the Fisher-Yates shuffle algorithm. 
A 128-bits key allows for list permutation of up to 34 positions (34 prediction probabilities) in a secure manner.

\subsubsection{Indistinguishability}
\label{sec:indistinguishability}

Outputs $B_\mathcal{V}(x)$ must be indistinguishable from $F_\mathcal{V}(x)$ \ref{req:indifferentiability}. This requirement is partially addressed by our assumption that \adv has no expectation regarding predictions obtained from $F_\mathcal{V}$ (Sect.~\ref{sec:assumption}).
Nevertheless, our watermarking function $W_\mathcal{V}$ is configured by a hash of the input $x$.
A subtle modification $\delta$ to $x$ produces a different hash and consequently, a different result $W_\mathcal{V}(x) \neq W_\mathcal{V}(x+\delta)$. If \adv receives different predictions for $x$ and $x + \delta$ for a small $\delta$, it can discard both $x$ and $x + \delta$ from its training set to avoid the watermark. 

Therefore we assume that \adv expects two similar inputs $x$ and $x+\delta$ to have similar predictions $F_\mathcal{V}(x) \sim F_\mathcal{V}(x+\delta)$ when $\delta$ is small.
To enhance indistinguishability, the decision of $W_\mathcal{V}$ must be smoothened to return the same result $W_\mathcal{V}(x) = W_\mathcal{V}(x+ \delta)$ and $B_\mathcal{V}(x) = B_\mathcal{V}(x+ \delta)$.
This can be achieved using a mapping function $M_\mathcal{V}: \mathbb{R}^n \rightarrow \mathbb{R}^p$ that projects $x$ to a space where $M_\mathcal{V}(x) = M_\mathcal{V}(x+\delta)$ for a small $\delta$.
$M_\mathcal{V}(x)$ is only used as the new input to our hash function such that $\text{HMAC}(K_w,M_\mathcal{V}(x)) = \text{HMAC}(K_w,M_\mathcal{V}(x+\delta))$.
$M_\mathcal{V}(x)$ smoothens the decision of $W_\mathcal{V}$ and ensures that permutations performed in $B_\mathcal{V}$ are the same for similar inputs ($\pi$ is keyed by the hash result).

$M_\mathcal{V}$ could be implemented as an autoencoder which projects inputs $x$ to a latent space $\mathbb{R}^p$ of lower dimension $p < n$, discarding small perturbations~\cite{meng2017magnet}.
$M_\mathcal{V}$ could also be a masking and binning function~\cite{cohen2019certified} removing large modifications of a single pixel value (with masking) and small modifications of a large number of pixels (with binning).
We evaluated one particular implementation of $M_\mathcal{V}$: use the embedding obtained from a layer in the middle of $F_\mathcal{V}$.
This is similar to using an autoencoder but it does not require training additional models.
The obtained embedding would also be unknown to \adv since it does not know $F_\mathcal{V}$ (target of extraction attack).

For each input $x$, we obtain its latent representation $L_\mathcal{V}$ based on $F_\mathcal{V}$ as $x_L = L_\mathcal{V} (F_\mathcal{V}, x)$.
This ensures that as long as $F_\mathcal{V}$'s prediction is resilient to perceptual modifications (e.g. translation, illumination), so is $M_\mathcal{V}$.
Next, we smoothen $x_L$ by binarizing it based on the median value of each of its features.
The median of each feature value is obtained by querying $F_\mathcal{V}$ using \victim's training set, recording corresponding $x_L$ and taking the median. Using 5000 samples and their intermediate representation of length 100, we get 100 feature vectors of length 5000 and thus, 100 median values.
We evaluate $M_\mathcal{V}$ in Sect.~\ref{sec:perfect-mapping}.

\subsection{Watermark embedding}

\adv uses the set of inputs $D_\mathcal{A}$ and the corresponding predictions returned by \ourname-protected prediction API of $F_\mathcal{V}$ to train $F_\mathcal{A}$.
Approximately $\lfloor r_w \times \vert D_\mathcal{A} \vert \rfloor$ samples from $D_\mathcal{A}$ constitute the trigger set $T_\mathcal{A}$ consisting of incorrect predictions $B_\mathcal{V}(x)$.
Given that $F_\mathcal{A}$ has enough capacity (large enough number of parameters), it will be able to remember a certain amount of training data having arbitrarily incorrect labels~\cite{zhang2016understanding}. This phenomenon is called overfitting and it can be prevented using regularization~\cite{bishop2006pattern}. But it is not effective for DNNs with a large capacity~\cite{zhang2016understanding}.
This is the rationale for the existence of DNN backdoors~\cite{liu2018trojaning} and for \ourname.
We expect our watermark $(T_\mathcal{A},\hat{B}_{\mathcal{V}}(T_\mathcal{A}))$ to be embedded as a backdoor in $F_\mathcal{A}$ as a natural effect of training a model $F_\mathcal{A}$ with high capacity. If the watermark is not embedded, we expect $F_\mathcal{A}$'s accuracy on the primary task to be too low to make it usable (\ref{req:unremovability}).

Different adversaries $\mathcal{A}_i$ will have different datasets $D_{\mathcal{A}_i}$. Consequently, the trigger sets $T_{\mathcal{A}_i}$ selected by \ourname will also be different. Different surrogate models $F_{\mathcal{A}_i}$ will embed distinctive watermarks. 
Each watermark thus links to the \client identifier. \ourname meets the linkability requirement~\ref{req:linkability}.

\subsection{Watermark verification}
\label{sec:own-proof}

We present the \textit{verify} function used by \judge to prove a model $F'$ is a surrogate of $F$.
\textit{Verify} tests if a given watermark $(T, \hat{B}(T))$ is embedded in a model $F'$ suspected to be a surrogate of $F$.
We first define $L(T,\hat{B}(T),F')$ that computes the ratio of different results between the backdoor function $\hat{B}(x)$ and the suspected surrogate model $\hat{F}'(x)$ for all inputs in the trigger set.

\begin{equation}
\label{eq:verification}
    L(T,\hat{B}(T),F') = \dfrac{1}{\vert T \vert} \sum_{x \in T} (\hat{F}'(x) \neq \hat{B}(x))
\end{equation}


The watermark verification succeeds, i.e., \textit{verify} returns ``true'',  if and only if $L(T,\hat{B}(T),F') < e$, where $e$ is a tolerated error rate that must be defined.
This means we must have at most $\lfloor e \times \vert T \vert \rfloor$ samples where $\hat{B}(x)$ and $\hat{F}'(x)$ differ in order to declare $F'$ is a surrogate of $F$.
The choice for the value of $e$ is a trade-off between correctness and completeness for watermark verification (reliability~\ref{req:non-trivial}).
Assume we want to use a pre-generated watermark $(T,\hat{B}(T))$ to verify if an arbitrary model $F'$ is a surrogate.
For simplicity, we assume a uniform probability of matching the prediction of a watermarked input $P(\hat{B}(x) = \hat{F}'(x)) = 1/m$, where $m$ is the number of classes of $F'$.
The probability for trivial watermark verification success, given a trigger set of size $\vert T \vert$ and an error rate $e$, can be computed using the cumulative binomial distribution function as follows.

\begin{equation}
\label{eq:proof}
    P(L < e) = \sum_{i=0}^{\lfloor e \times \vert T \vert \rfloor} \binom{\vert T \vert}{i} \times \left( \dfrac{m-1}{m} \right) ^{i} \times \left( \dfrac{1}{m} \right) ^{\vert T \vert-i}
\end{equation}

\begin{figure}[th]
                \centering
                \includegraphics[width=0.95\columnwidth]{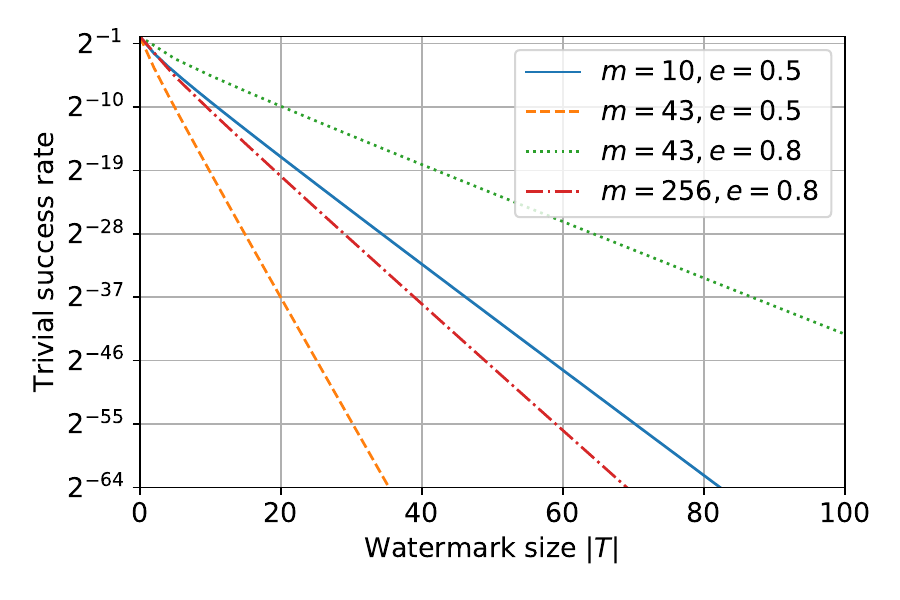}
                \caption{Resilience to trivial watermark verification vs. size of the watermark for different tolerated error rates $e$ and models with different number of classes $m$. The success rate for trivial watermark verification decreases exponentially with the watermark size.}
                \label{fig:wm-size}
\end{figure}

This probability is the average success rate of \adv wanting to frame \victim for model stealing using an arbitrary watermark.
Figure~\ref{fig:wm-size} depicts the decrease of this success rate as we increase the watermark size.
We see that the verification function can accommodate a large error rate ($e > 0.5$) while preventing trivial success in verification using a small watermark ($\vert T \vert \approx 50$).
The error rate $e$ must be defined proportionally to the number of classes $m$. Large error rates can be used for models with a large number of classes.
For instance, we can set $e = 0.8$ for a model with $m = 256$ classes, limiting the adversary success rate to less than $2^{-64}$ for a watermark of size 70.

The success rate in trivial verification is the complement of the confidence for reliable watermark verification, and for reliable demonstration of ownership by transition $1 - P(L < e)$.
The choice of $e$ defines the minimum watermark size given a targeted confidence. Recall that this size must also be small to ensure utility of the model to protect \ref{req:qos}.
The tolerated error must necessarily be lower than the probability of random class match: $e < (1-m)/m$.
Also, $e$ must be larger than $\epsilon$ where $Acc(F_\mathcal{A}) = 1 - \epsilon$ is the accuracy of the watermarked surrogate model $F_\mathcal{A}$ on the trigger set.

The success of watermark verification is not sufficient to declare ownership of a surrogate model $F'$.
\adv can increase its success in trivial watermark verification from random using several means.
For instance, knowing $F$ and $F'$, \adv can find inputs $x$ for which $F(x) \neq F'(x)$ and use pairs $(x, F'(x))$ as a watermark that would successfully pass watermark verification.
Thus demonstrating ownership requires a careful process to ensure that the probability for matching an incorrect prediction class remains random, ensuring that the probability for trivial watermark verification follows Eq.~\ref{eq:proof}.


%
%

\subsection{Demonstrating ownership}
\label{sec:registration-verification}

We present the process for a model owner \victim to demonstrate ownership of a surrogate model watermarked by \ourname. It only requires the suspected surrogate model $F_\mathcal{A}$ to expose a prediction API.
This process uses a judge $\mathcal{J}$ who is trusted to (a) ensure confidentiality of all data submitted as input to the process and (b) correctly execute and report the results of the specified \emph{verify}.
It also uses a time-stamped public bulletin board, e.g., a blockchain, in which information can be published to provide proof of anteriority.
$\mathcal{J}$ can be implemented using an trusted execution environment (TEE)~\cite{DBLP:journals/ieeesp/EkbergKA14}.

\subsubsection{Watermark registration}

\victim publishes cryptographic commitments of the following elements in the public bulletin board:
\begin{itemize}
    \item the model $F_\mathcal{V}$.
    \item for each API client $i$, one registered watermark $(T_{\mathcal{A}_i},\hat{B}_\mathcal{V}(T_{\mathcal{A}_i}))$.
\end{itemize}

The commitment can be instantiated using a cryptographic hash function $H(\:)$, e.g., SHA-3. Each watermark should be linked to the corresponding model, e.g., by associating $H(F_\mathcal{V})$ with each registered watermark.

Several updated versions of the registered watermark can be published for each \client, as they make more queries to the prediction API and their watermarks grow. The verification of any one of these watermarks is sufficient to demonstrate ownership of the model.
We define the following rules for reliable demonstration of ownership~\ref{req:non-trivial} that prevents ownership piracy~\ref{req:non-piracy}:
\begin{itemize}
    \item $\left( H(T_{\mathcal{A}_i},\hat{B}_\mathcal{V}(T_{\mathcal{A}_i})) , H(F_\mathcal{V}) \right)$ is valid only if published later than $H(F_\mathcal{V})$.
    \item \adv can refute $F_\mathcal{A}$ is a surrogate model only if $H(F_\mathcal{A})$ has been published.
    \item $\left( H(T_{\mathcal{A}_i},\hat{B}_\mathcal{V}(T_{\mathcal{A}_i})) , H(F_\mathcal{V}) \right)$ can only demonstrate that $F_\mathcal{A}$ is a surrogate of $F_\mathcal{V}$ if $H(F_\mathcal{A})$ is published later than $H(F_\mathcal{V})$ (or not published at all).
    \item in case of contention, the model having its commitment first published is deemed to be the original.
\end{itemize}

\subsubsection{Verification process}

When \victim suspects a model $F_\mathcal{A}$ is a surrogate of $F_\mathcal{V}$ trained by an \client $i$, it provides a pointer to the prediction API of $F_\mathcal{A}$ to \judge.
It also provides the following secret information using a confidential communication channel: the \client $i$ watermark $(T_{\mathcal{A}_i},\hat{B}_\mathcal{V}(T_{\mathcal{A}_i}))$ and $F_\mathcal{V}$.
\judge does the following to check if $F_\mathcal{A}$ is a surrogate of $F_\mathcal{V}$. If any step fails, the ownership of $F_\mathcal{A}$ is not considered to have been demonstrated. If all succeed, \judge gives the verdict that $F_\mathcal{A}$ is a surrogate of $F_\mathcal{V}$.
\begin{enumerate}
    \item compute $H(T_{\mathcal{A}_i},\hat{B}_\mathcal{V}(T_{\mathcal{A}_i}))$ and use it as a pointer to retrieve the registered watermark $\left( H(T_{\mathcal{A}_i},\hat{B}_\mathcal{V}(T_{\mathcal{A}_i})) , H(F_\mathcal{V}') \right)$ from the public bulletin.
    \item compute  $H(F_\mathcal{V})$ and verify $H(F_\mathcal{V}) = H(F_\mathcal{V}')$, where $H(F_\mathcal{V}')$ is extracted from the registered watermark.
    \item retrieve $H(F_\mathcal{V})$ from the public bulletin and verify it was published before $\left( H(T_{\mathcal{A}_i},\hat{B}_\mathcal{V}(T_{\mathcal{A}_i})) , H(F_\mathcal{V}') \right)$.  
    \item query $T_{\mathcal{A}_i}$ to $F_\mathcal{A}$'s prediction API and verify that
    \begin{sloppypar}\noindent
    $L(T_{\mathcal{A}_i},\hat{B}_\mathcal{V}(T_{\mathcal{A}_i}),F_\mathcal{A}) < e$.
    \end{sloppypar}
    \item input $T_{\mathcal{A}_i}$ to $F_\mathcal{V}$ and verify $\hat{B}_\mathcal{V}(x) \neq \hat{F}_\mathcal{V}(x), \forall x \in T_{\mathcal{A}_i}$.
\end{enumerate}

If $F_\mathcal{A}$'s owner (\adv) wants to contest the verdict, it must provide the original model $F'_\mathcal{A}$ to \judge using a confidential communication channel.
\judge assesses that the provided model and the API model are the same $F_\mathcal{A} = F'_\mathcal{A}$ by verifying $F_\mathcal{A}(x) = F'_\mathcal{A}(x), \forall x \in T_{\mathcal{A}_i}$.
Then, \judge computes $H(F'_\mathcal{A})$ and retrieves it from the public bulletin.
If $H(F'_\mathcal{A})$ was published before $H(F_\mathcal{V})$, \judge concludes that $F'_\mathcal{A} = F_\mathcal{A}$ is an original model.

\section{Experimental setup}
\label{sec:exp-setup}

\subsection{Datasets and Models}\label{datasets-and-models}

\subsubsection{Datasets}

We evaluate \ourname using four image recognition datasets that were used in prior work to evaluate DNN extraction attacks.
MNIST~\cite{lecun2010mnist} (60,000 train and 10,000 test samples, 10 classes) and GTSRB~\cite{stallkamp2011german} (39,209	train and 12,630 test samples, 43 classes) are respectively a handwritten-digit and traffic-sign dataset used to showcase the extraction of low capacity DNN models~\cite{juuti2019prada, papernot2017practical}.
CIFAR10~\cite{krizhevsky2009cifar10} (50,000 train and 10,000 test samples, 10 classes) and Caltech256~\cite{griffin2007caltech} (23,703 train and 6,904 test samples, 256 classes) contain images depicting miscellaneous objects that were used to showcase the extraction of high capacity DNN models~\cite{orekondy2018knockoff,correia2018copycat}.

We also selected a random subset of 100,000 samples from ImageNet dataset~\cite{deng2009imagenet} (1000 classes), which contains images of natural and man-made objects.
We use it to evaluate the embedding of different types of watermarks and to perform a model extraction attack that requires such samples~\cite{orekondy2018knockoff}.

\subsubsection{Models}
\label{sec:models-pres}

We select two kinds of DNN models to evaluate the embedding of a watermark: low-capacity models having less than 10M parameters, and high-capacity models having over 20M parameters.
These models are presented in Table~\ref{tab:models}.

\begin{table}[htb]
\begin{center}
	\caption{DNN models used to evaluate \ourname. Number of training epochs and base test accuracy.}
	\label{tab:models}
	\begin{tabular}{lccccc} \hline
		Model				& Input size 			& $m$		& Param.		& Epochs	& $Acc_{test}$ \\ \hline
		MNIST-3L			& 28x28x1			& 10				& 62,346 			& 10			& 98.6 \\
		MNIST-5L			& 28x28x1			& 10				& 683,522 			& 10			& 99.1 \\
		GTSRB-5L			& 32x32x3			& 43				& 669,123 			& 50			& 91.7 \\
		CIFAR10-9L		& 32x32x3			& 10				& $\sim$ 6 M 		& 100			& 84.6 \\ \hline
		GTSRB-RN34  	& 224x224x3		& 43				& $\sim$ 21 M 	& 250			& 98.1 \\
		CIFAR10-RN34 & 224x224x3		& 10				& $\sim$ 21 M		& 250			& 94.7 \\
		Caltech-RN34	& 224x224x3		& 256			& $\sim$ 21 M		& 250			& 74.4 \\
	\end{tabular}
\end{center}
\end{table}

In order to accurately reconstruct model extraction attacks, we use the same model architectures and training process as in~\cite{juuti2019prada} for low-capacity models and as in~\cite{orekondy2018knockoff} for high-capacity models.
Similarly to prior work~\cite{orekondy2018knockoff}, we use ResNet34~\cite{he2016deep} architecture pre-trained on ImageNet as a basis for high-capacity models.
We fine-tuned Caltech-RN34, GTSRB-RN34 and CIFAR10-RN34 models using Caltech256, GTSRB and CIFAR10 datasets respectively~\footnote{We chose to reproduce only the Caltech-RN34 experiment from~\cite{orekondy2018knockoff} because of its best performance. We used CIFAR10 and GTSRB to conduct supplementary experiments with high capacity models as they allow us to juxtapose results of experiments with low and high capacity models on the same datasets.}.
We also trained DenseNet121~\cite{huang2017densely} models to perform additional experiments due to the absence of dropout layers in ResNet34 models.
All models were trained using Adam optimizer with learning rate of 0.001 that was decreased over time to 0.0005 (after 100 epochs for ResNet34 models and half-way for the other), except for Caltech-RN34. For Caltech-RN34, we used SGD optimizer with an initial learning rate of 0.1 that was decreased by a factor of 10 every 60 epochs over 250 epochs. We used a batch size of 16 for fine-tuning ResNet34 and DenseNet121 based models.

\subsection{Watermarking Procedure}\label{watermarking-procedure}

Inputs from \adv's dataset $D_\mathcal{A}$ are submitted to the \ourname-enhanced prediction API of $F_\mathcal{V}$ which returns correct $F_\mathcal{V}(x)$ or incorrect predictions $B_\mathcal{V}(x)$ according to the result of the watermarking function $W_\mathcal{V}(x)$.
For the experiments in Section~\ref{sec:perfect-mapping} (evaluating the effectiveness of the mapping function $M_\mathcal{V}$), we use the embedding from $F_\mathcal{V}$ as $M_\mathcal{V}$. Experiments in Section~\ref{sec:unremovability} and Section~\ref{sec:eval-stealing} do not depend on the choice of $M_\mathcal{V}$. Therefore, for the sake of simplicity, we use the identity function as $M_\mathcal{V}$ in these experiments.

We simulate \adv who uses the whole set $D_\mathcal{A}$, which includes $\vert T_\mathcal{A} \vert$ samples with incorrect labels, to train its surrogate model $F_\mathcal{A}$.
\adv trains $F_\mathcal{A}$ without being aware of the watermarked samples in $D_\mathcal{A}$.

\subsection{Evaluation Metrics}\label{sec:metrics}

We use two metrics to evaluate the success of \adv's goal and \victim's goal respectively.
\adv's goal is to train a surrogate model $F_\mathcal{A}$ that has maximum accuracy on $F_\mathcal{V}$'s primary classification task. We evaluate this by computing the \textit{test accuracy} of the surrogate model $Acc_{test}(F_\mathcal{A})$ on the test set $Test$ of each dataset.

\begin{equation}
    Acc_{test}(F_\mathcal{A}) = \frac{1}{|Test|}\sum_{x_i \in Test}
    \begin{cases}
      1, & \text{if}\ \hat{F}_\mathcal{A}(x_i) = O_f(x_i)   \\
      0, & \text{otherwise}
    \end{cases}
\end{equation}

\victim's goal is to maximize the embedding of the watermark in any surrogate model built from responses from $F_\mathcal{V}$ such that its surrogacy can be reliably demonstrated. We evaluate this by computing the \textit{watermark accuracy} of the surrogate model $Acc_{wm}(F_\mathcal{A})$ on the trigger set $T_\mathcal{A}$ of watermarked inputs.

\begin{equation}
    Acc_{wm}(F_\mathcal{A}) = \frac{1}{|T_\mathcal{A}|}\sum_{x_i \in T_\mathcal{A}}
    \begin{cases}
      1, & \text{if}\ \hat{F}_\mathcal{A}(x_i) = \hat{B}_\mathcal{V}(x_i)   \\
      0, & \text{otherwise}
    \end{cases}
\end{equation}

\ourname aims to maximize $Acc_{wm}(F_\mathcal{A})$ regardless of
 $Acc_{test}(F_\mathcal{A})$. \adv aims to maximize  $Acc_{test}(F_\mathcal{A})$ while minimizing $Acc_{wm}(F_\mathcal{A})$.
In our experiments, we calculate both metrics every 5 epochs in order to evaluate their progress during the training process.

\section{Robustness of watermarking}
\label{sec:eval-perfect}

We assess \adv's ability to prevent the embedding of a watermark in a surrogate model, i.e., to violate the unremovability requirement \ref{req:unremovability}.
Prior work evaluated unremovability \textit{after} a watermarked model is trained showing that backoor-based watermarks are resilient to model pruning and adversarial fine tuning~\cite{adi2018turning,merrer2017adversarial,zhang2018protecting}.
\ourname also embeds backdoor-based watermarks resilient to removal using post-training manipulations.
Thus, we focus on adversarial manipulations \textit{during} training by evaluating several solutions that could prevent watermark embedding.
We then evaluate the ability for \adv to identify watermarked inputs using the trained surrogate model, i.e., to violate the indistinguishability requirement~\ref{req:indifferentiability}.

We take an ideal model extraction attack scenario where $\hat{F}_\mathcal{V} = O_f$ is a perfect oracle.
\adv has access to a large dataset $D_\mathcal{A}$ of natural samples from the same distribution as \victim training data: we use the whole training set from each dataset (Sect.~\ref{datasets-and-models}) for $D_\mathcal{A}$.
We use a large watermark of fixed size $\vert T_\mathcal{A} \vert = 250$ in all following experiments. Embedding a large watermark is challenging since the model must learn many isolated errors (mislabeled inputs). We take $\vert T_\mathcal{A} \vert = 250$ as an upper bound to the watermark size and a worst case scenario for \ourname watermark embedding.

\subsection{Unremovability of watermark during training}
\label{sec:unremovability}

We evaluate the impact of two parameters on embedding a watermark during DNN training.
The first parameter is the capacity of  $F_\mathcal{A}$. \adv can limit this capacity such that the model could only learn the primary classification task and cannot learn the watermark.
The second parameter is the use of regularization. Regularization accommodates classification errors on the training data, which is considered as noise. The watermark consists of incorrectly labeled inputs which can potentially be discarded using regularization.

We evaluate the impact of model capacity and regularization on watermark accuracy $Acc_{wm}$ and test accuracy $Acc_{test}$ of $F_\mathcal{A}$.
We trained several surrogate models having low and high capacity. $T_\mathcal{A}$ was randomly selected from the respective training sets.
We used plain training and two regularization methods, namely weight decay~\cite{krogh1992simple} with decaying factor $\lambda$ and dropout (DO=X)~\cite{srivastava2014dropout} with probability X=$\left\lbrace 0.3,0.5\right\rbrace$. We selected $\lambda$ values optimal for \adv: such that they maximize the difference $Acc_{test} - Acc_{wm}$.

Table~\ref{tab:embedding-low} and~\ref{tab:embedding-high} present the results of this experiment for DNN models with low and high capacity respectively.
We report $Acc_{wm}$ and $Acc_{test}$ results at three training stages providing (1) best watermark accuracy (best for \victim), (2) best test accuracy (best for \adv) and (3) when training is completed.
Overall, we observe that  $Acc_{test}$ and $Acc_{wm}$ are high for most settings.
Using plain training, $Acc_{wm}$ is mostly higher than $Acc_{test}$ and often close to 100\%. The ownership of all these surrogate models can be reliably demonstrated using a low tolerated error rate, e.g., $e=0.3$.

\setlength\tabcolsep{3.5pt}
\begin{table}
    \caption{Impact of regularization -- dropout (DO) and weight decay ($\lambda$) -- on test ($test$) and watermark accuracy ($wm$) of surrogate models $F_\mathcal{A}$. We report results at the training epoch (ep.) reaching best $Acc_{wm}$ (optimal for \victim), best $Acc_{test}$  (optimal for \adv) and when training is over (Final). Purple (underline) results highlight low $Acc_{wm}$ and $Acc_{test}$: $F_\mathcal{A}$ is unusable. Red (dashed underline) results highlight low $Acc_{wm}$ ($<50\%$) while $Acc_{test}$ remains significantly high (decrease $<10pp$): \victim may fail to prove ownership of $F_\mathcal{A}$.}
    \label{tab:embedding}

\begin{subtable}{\linewidth}\centering
    \caption{Low capacity models. 500 training epochs.}\label{tab:embedding-low}
    \resizebox{1.\columnwidth}!{
{\begin{tabular}{llll|lll|ll} \hline
                                                & \multicolumn{3}{c|}{Best $Acc_{wm}$} 		    & \multicolumn{3}{c|}{Best $Acc_{test}$}	        & \multicolumn{2}{c}{Final}\\
        Model									& $wm$			    & $test$		    & ep.	& $wm$			    & $test$		    & ep. 		& $wm$			    & $test$\\ \hline
        MNIST-3L 								& \accallfail{14\%} & \accallfail{11\%}	& -		& \accallfail{14\%}	& \accallfail{11\%}	& -			& \accallfail{14\%}	& \accallfail{11\%}\\
        MNIST-3L (DO=0.3)						& \accallfail{12\%}	& \accallfail{11\%}	& -		& \accallfail{12\%}	& \accallfail{11\%}	& -			& \accallfail{12\%}	& \accallfail{11\%}\\
        MNIST-3L (DO=0.5)						& \accallfail{13\%}	& \accallfail{11\%}	& -		& \accallfail{13\%}	& \accallfail{11\%}	& -			& \accallfail{13\%}	& \accallfail{11\%}\\
        MNIST-3L ($\lambda=5e^{-6}$)			& 99\%			    & 89\%			    & 210	& 98\%			    & 96\%			    & 290		& 97\%			    & 96\%\\ \hline
        MNIST-5L								& 99\%			    & 96\%			    & 120	& 98\%			    & 97\%			    & 170		& 98\%			    & 97\%\\
        MNIST-5L (DO=0.3)						& \accallfail{13\%} & \accallfail{16\%}	& 50	& \accallfail{11\%}	& \accallfail{51\%}	& 30		& \accallfail{12\%}	& \accallfail{11\%}\\
        MNIST-5L (DO=0.5)						& \accallfail{13\%} & \accallfail{17\%}	& 15	& \accallfail{9\%}	& \accallfail{53\%} & 50		& \accallfail{11\%}	& \accallfail{13\%}\\
        MNIST-5L ($\lambda=5e^{-6}$)			& 99\% 			    & 88\% 			    & 215 	& 99\% 			    & 94\% 			    & 365 		& 98\%			    & 93\%\\ \hline
        GTSRB-5L								& 97\% 			    & 88\% 			    & 160 	& 95\%			    & 89\%			    & 190		& 97\%			    & 88\%\\
        GTSRB-5L (DO=0.3)						& 99\% 			    & 88\% 			    & 135 	& 98\%			    & 90\%			    & 220		& 98\%			    & 88\%\\
        GTSRB-5L (DO=0.5)						& 98\% 			    & 89\% 			    & 105 	& 98\%			    & 90\%			    & 200		& 98\%			    & 89\%\\
        GTSRB-5L ($\lambda=5e^{-6}$)	        & \accallfail{28\%} & \accallfail{55\%} & 410 	& \accwmfail{17\%} 	& \accwmfail{71\%}	& 105		& \accwmfail{25\%}	& \accwmfail{79\%}\\ \hline
        CIFAR10-9L								& 93\% 			    & 78\% 			    & 110 	& 92\%			    & 79\%			    & 105		& 73\%			    & 76\%\\
        CIFAR10-9L (DO=0.3)						& 40\% 			    & 75\% 			    & 125 	& \accwmfail{35\%}	& \accwmfail{75\%}	& 90		& \accwmfail{25\%}	& \accwmfail{70\%}\\
        CIFAR10-9L (DO=0.5)						& 45\% 			    & 71\% 			    & 240 	& \accwmfail{25\%} 	& \accwmfail{77\%}  & 90		& \accwmfail{25\%}	& \accwmfail{75\%}\\
        CIFAR10-9L ($\lambda=3e^{-4}$)			& \accallfail{32\%} & \accallfail{72\%} & 235 	& \accwmfail{32\%} 	& \accwmfail{72\%}	& 235 		& \accallfail{25\%}	& \accallfail{47\%}\\ \hline
    \end{tabular}}}
\end{subtable}\\
\vspace*{1 em}

\begin{subtable}{\linewidth}\centering
    \caption{High capacity models. 250 training epochs.}\label{tab:embedding-high}
    \resizebox{1.\columnwidth}!{
{\begin{tabular}{llll|lll|ll} \hline
                                         & \multicolumn{3}{c|}{Best $Acc_{wm}$}	        & \multicolumn{3}{c|}{Best $Acc_{test}$}	        & \multicolumn{2}{c}{Final}\\
        Model 						 	 & $wm$ 			 & $test$ 		     & ep.	& $wm$ 			    & $test$ 		    & ep.		& $wm$			    & $test$\\ \hline
        GTSRB-RN34 					 	 & 83\%			     & 97\% 			 & 245	& 70\% 			    & 98\% 			    & 105 		& 84\%			    & 97\%\\
        GTSRB-DN121 (DO=0.3) 			 & 98\%			     & 89\% 			 & 240	& 98\% 			    & 89\% 			    & 240 		& 95\%			    & 86\%\\
        GTSRB-DN121 (DO=0.5) 			 & 99\%			     & 92\%			     & 235 	& 98\% 			    & 93\% 			    & 245		& 98\% 			    & 93\%\\
        GTSRB-RN34 ($\lambda=e^{-5}$) 	 & 87\%			     & 92\%			     & 200 	& 87\%			    & 92\% 			    & 200 		& 73\% 			    & 77\%\\ \hline
        CIFAR10-RN34 					 & 99\% 			 & 89\% 			 & 110 	& 99\% 			    & 90\% 			    & 240 		& 98\%			    & 89\%\\
        CIFAR10-DN121 (DO=0.3)			 & 99\% 			 & 88\% 			 & 160 	& 98\% 			    & 88\% 			    & 210 		& 97\% 			    & 86\%\\
        CIFAR10-DN121 (DO=0.5)			 & 99\% 			 & 85\% 			 & 130 	& 97\% 			    & 88\% 			    & 220 		& 98\% 			    & 87\%\\
        CIFAR10-RN34 ($\lambda=e^{-5}$)	 & 100\% 		     & 80\% 			 & 10 	& 100\% 		    & 89\% 			    & 160 		& 97\% 			    & 81\%\\ \hline
        Caltech-RN34					 & 97\% 			 & 69\% 			 & 110 	& 93\% 			    & 73\% 			    & 160 		& 94\% 			    & 73\%\\
        Caltech-DN121 (DO=0.3)			 & \accallfail{48\%} & \accallfail{44\%} & 110 	& \accwmfail{36\%} 	& \accwmfail{59\%} 	& 155 		& \accwmfail{32\%} 	& \accwmfail{57\%}\\
        Caltech-DN121 (DO=0.5)			 & \accallfail{35\%} & \accallfail{30\%} & 115 	& \accallfail{22\%} & \accallfail{49\%} & 185 		& \accallfail{21\%} & \accallfail{49\%}\\
        Caltech-RN34 ($\lambda=3e^{-4}$) & 89\% 		     & 67\% 			 & 100 	& 69\% 			    & 68\% 			    & 60 		& 76\% 			    & 68\%\\ \hline
    \end{tabular}}}

    \end{subtable}

\end{table}

\textbf{Model capacity.}
High-capacity models can provide higher watermark and test accuracy than low-capacity models as highlighted by comparing results for GTSRB and CIFAR10 in both tables.
While $Acc_{wm}$ is low for some low-capacity models, e.g., MNIST-3L, MNIST-5L (DO), their test accuracy is similarly low and close to random $Acc_{wm} \sim Acc_{test} \sim 10\%$.
This shows that reducing the model capacity can prevent the embedding of the watermark.
However, decreasing $Acc_{wm}$ to a level where it cannot be used to reliably prove ownership makes $F_\mathcal{A}$ unusable.
$Acc_{wm}$ and $Acc_{test}$ are closely tied when manipulating the model capacity and thus this is not a useful strategy to circumvent \ourname.



\textbf{Regularization.}
Regularization is useful for decreasing the watermark accuracy in a few cases.
Weight decay is useful for low-capacity GTSRB-5L and CIFAR10-9L models.
Dropout is useful for low-capacity MNIST-5L and CIFAR10-9L models, and for high-capacity Caltech-DN121 model.
Dropout completely prevents the embedding of the watermark into MNIST-5L model as depicted by $Acc_{wm} \sim 10\%$.
However, $Acc_{test}$ is also significantly reduced, by 50\% at best, making $F_\mathcal{A}$ potentially unusable.
\techreport{Figure~\ref{fig:accuracy-evolution} (MNIST-5L (DO=0.3)) further highlights that the maximal test accuracy of 51\% is difficult to obtain since $Acc_{test}$ is very unstable along the training process, changing abruptly from 10\% to 50\% while the training loss remains constantly low.}
In all remaining cases, $Acc_{wm}$ is reduced down to 20-35\%, while preserving high test accuracy similar to models trained with non-watermarked datasets.
While $Acc_{wm}$ is low, the watermark can still successfully demonstrate ownership by increasing the tolerated error rate to, e.g., $e = 0.8 > 1- Acc_{wm}$.
Considering the large watermark size of 250, this demonstration would still be reliable despite the high tolerated error rate as evaluated in Sect.~\ref{sec:own-proof}.

It is worth noting that no regularization method is effective at removing the watermark from high capacity GTSRB-RN34 and CIFAR10-RN34 models. The likely reason is that ResNet34 architecture has significant overcapacity for the primary task of classifying these datasets.
Regularization cannot limit this capacity to an extent where the watermark would not be embedded.
This means \adv needs sufficient knowledge of $F_\mathcal{V}$ to select an appropriate model architecture for $F_\mathcal{A}$. It must have sufficient capacity to learn the primary classification task of the victim model while preventing watermark embedding.
In model extraction attacks, \adv has black-box access to $F_\mathcal{V}$, which forces to use $F_\mathcal{A}$ with sufficient capacity to maximize the attack success~\cite{orekondy2018knockoff}. In this setting, regularization is not useful to circumvent \ourname.

Finally, while regularization can be useful, \adv needs relevant test data and ground truth to optimize the regularization parameters (e.g., decaying factor $\lambda$).
\techreport{Also, we observed in Fig.~\ref{fig:accuracy-evolution} that test accuracy is very unstable while the training loss remains constantly low when using regularization. \adv needs additional labeled test data to apply early stopping~\cite{caruana2001overfitting} of training at the optimal epoch providing maximal test accuracy.}
In all extraction attacks~\cite{tramer2016stealing,  juuti2019prada, orekondy2018knockoff, papernot2017practical, correia2018copycat,pal2020activethief} the availability of relevant data is the main limitation. All this data is typically used for training the surrogate model and none is used for test purposes, which prevents optimization of regularization parameters and early stopping.

\techreport{A third parameter that may impact watermark embedding is the distribution of watermarked inputs. In contrast to prior DNN watermarking solutions\cite{merrer2017adversarial,adi2018turning,zhang2018protecting}, \ourname uses watermarked inputs from the same distribution as the training data. The watermark distribution cannot be controlled by \adv.}

\subsection{Mapping Function}
\label{sec:perfect-mapping}

\adv can try to identify watermarked inputs and remove them from $D_\mathcal{A}$ prior to training in order to prevent watermark embedding. Because \ourname relies on a hash to decide if an input is watermarked, \adv can query multiple perturbed versions of inputs in $D_\mathcal{A}$ and discard those that return different predictions.
The mapping function $M_\mathcal{V}$ presented in Sect.~\ref{sec:indistinguishability} is meant to prevent this evasion.
We evaluate the effectiveness of $M_\mathcal{V}$ by querying 10 perturbed versions of each of the 10,000 samples in $D_\mathcal{A}$, which includes $|T_\mathcal{A}| = 121$. For each query, we check whether they get consistent mapping $M_\mathcal{V}$ and classification $\hat{F}_\mathcal{V}$.
Table~\ref{tab:mapping} reports the results of this experiment for various perturbation size $\delta$ for the MNIST dataset.
We distinguish cases where
\circled{1}
$\hat{F}_\mathcal{V}(x) = \hat{F}_\mathcal{V}(x + \delta)$ (same $\hat{F}_\mathcal{V}$) or
$\hat{F}_\mathcal{V}(x) \neq \hat{F}_\mathcal{V}(x + \delta)$ (diff $\hat{F}_\mathcal{V}$);
\circled{2}
$M_\mathcal{V}(x) = M_\mathcal{V}(x + \delta)$ (same $M_\mathcal{V}$) or
$M_\mathcal{V}(x) \neq M_\mathcal{V}(x + \delta)$ (diff $M_\mathcal{V}$).
Same $\hat{F}_\mathcal{V}$ and same $M_\mathcal{V}$ means \adv keeps a watermarked sample in $D_\mathcal{A}$ ($M_\mathcal{V}$ succeeds).
Same $\hat{F}_\mathcal{V}$ and different $M_\mathcal{V}$ means \adv discards a watermarked sample from $D_\mathcal{A}$ ($M_\mathcal{V}$ fails).
Different $\hat{F}_\mathcal{V}$ means \adv wrongfully discards a sample from $D_\mathcal{A}$ regardless of $M_\mathcal{V}$ ($\delta$ is too large and changes $F_\mathcal{V}$'s prediction).
We see $M_\mathcal{V}$ succeeds to provide a consistent mapping in over 85\% cases for $\delta \leq 0.1$, meaning 85\% of the $T_\mathcal{A}$ is preserved in $D_\mathcal{A}$. As perturbations $\delta$ increase in size, $M_\mathcal{V}$ returns an increasing rate of inconsistent mapping, but this rate is similar to the one of changed predictions from $F_\mathcal{V}$. Thus, we conclude $M_\mathcal{V}$ is resilient to perturbations and \ourname can effectively watermark $F_\mathcal{A}$.


\begin{table}
    \caption{Resistance of $M_\mathcal{V}$ to perturbations $\delta$ of various size for the MNIST dataset.}
    \label{tab:mapping}
    \resizebox{1.\columnwidth}!{
    \begin{tabular}{l|c|c|c|c|c|c} \hline
                    & \multicolumn{3}{c|}{Entire  $D_\mathcal{A}$}                                  & \multicolumn{3}{c}{$T_\mathcal{A}$ only}\\
                    & \multicolumn{2}{c|}{same $\hat{F}_\mathcal{V}$} &	  diff $\hat{F}_\mathcal{V}$      & \multicolumn{2}{c|}{same $\hat{F}_\mathcal{V}$} &	  diff $\hat{F}_\mathcal{V}$  \\
        $\delta$  & same $M_\mathcal{V}$ & diff $M_\mathcal{V}$ &    & same $M_\mathcal{V}$ & diff $M_\mathcal{V}$ & \\ \hline
        0.2         & 99.30\%              & 0.44\%               & 0.26\%                  & 73.88\%              & 13.55\%              & 12.57\% \\
        0.1         & 99.63\%              & 0.24\%               & 0.13\%                  & 85.12\%              & 7.52\%               & 7.36\% \\
        0.09        & 99.64\%              & 0.22\%               & 0.14\%                  & 85.70\%              & 8.01\%               & 6.29\% \\
        0.075       & 99.71\%              & 0.19\%               & 0.10\%                  & 88.84\%              & 4.55\%               & 6.61\% \\
        0.05        & 99.81\%              & 0.12\%               & 0.07\%                  & 92.98\%              & 3.97\%               & 3.05\%
    \end{tabular}}
\end{table}

\section{Protecting against model extraction attacks}
\label{sec:eval-stealing}

We evaluate \ourname's effectiveness at watermarking surrogate DNN models constructed using two model extraction attacks: 1) the PRADA attack~\cite{juuti2019prada} achieves state-of-the-art performance in extracting low-capacity DNN models primarily using synthetic data and we launch it against MNIST-5L, GTSRB-5L and CIFAR10-9L;
2) the KnockOff attack~\cite{orekondy2018knockoff} extracts high-capacity DNN models using only natural data and we launch it against GTSRB-RN34, CIFAR10-RN34 and Caltech-RN34.
The test accuracy of each $F_\mathcal{A}$ extracted with these respective attacks is reported in Tab.~\ref{tab:stealing-complex}.

We demonstrate how to setup \ourname to protect a given victim model $F_\mathcal{V}$. We evaluate the successful embedding of watermarks in several surrogate models $F_\mathcal{A}$ as well as their utility considering a circumvention strategy.

\subsection{Effectiveness of DAWN}
\label{sec:dawn_settings}

\noindent\textbf{Watermarking decision:} \ourname degrades $F_\mathcal{V}$ utility by a factor equal to $r_w \times Acc(F_\mathcal{V})$ due to incorrect predictions for watermarked inputs.
The value of $r_w$ is specific to $F_\mathcal{V}$.
Given a desired level of confidence for reliable ownership demonstration equal to $1 - P(L < e)$ (cf. Eq.~\ref{eq:proof}), a tolerated error rate $e$ and the number of classes $m$ for $F_\mathcal{V}$, we can compute the minimum size for the watermark $\vert T_\mathcal{A} \vert$ using Eq.~\ref{eq:proof}.
Given that \victim can estimate the minimum number of queries $N$ required by \adv to train a usable surrogate model for $F_\mathcal{V}$, we can compute $r_w = N / \vert T_\mathcal{A} \vert$.
This ratio ensures that if \adv can successfully train a usable surrogate model $F_\mathcal{A}$, then $F_\mathcal{A}$ will embed a watermark large enough to reliably demonstrate its ownership .

The probability for successful trivial watermark verification $P(L < e)$ is valid for testing a single watermark. This probability increases by a factor equal to the number of tested watermarks.
\ourname creates and registers client-specific watermarks.
\victim must estimate the number of \client{s} to calculate the actual probability for trivial demonstration of ownership considering that all registered watermarks should be tested.
When verifying a watermark, the judge \judge counts the number of registered watermarks for $F_\mathcal{V}$ in the public bulletin.
\judge computes the real probability for successful trivial watermark verification accordingly and decides if a demonstration of ownership is reliable or not according to this final confidence.

\setlength\tabcolsep{4pt}
\begin{table}[htb]
\begin{center}
    \caption{Ratio of watermarked inputs $r_w$ required to protect six victim models $F_\mathcal{V}$ from extraction attack (PRADA for 3 first models / KnockOff for 3 last). Prediction API with 1M clients and targeted confidence for reliable demonstration of ownership  $= 1- 2^{-64}$. Number of attack queries ($N$) obtained from~\cite{juuti2019prada,orekondy2018knockoff} and used to compute the watermark size $\vert T_\mathcal{A} \vert$. $F_\mathcal{V}$ test accuracy decreases in a negligible manner ($r_w < 0.5\%$) that does not impact its utility.}
    \label{tab:WM_size}
    \resizebox{1.\columnwidth}!{
    \begin{tabular}{lcccccc} \hline
         Model	& classes & queries ($N$)	& $\vert T_\mathcal{A} \vert$	& $r_w (\%)$  & New $Acc(F_\mathcal{V})$  	\\ \hline
         MNIST-5L &	10  & 25,600		& 109	(0.1MB)  	& 0.426 & 98.7\% \\
         GTSRB-5L &	43  & 	25,520	& 47	 (0.4MB)  & 0.184 &  91.5\% \\
         CIFAR10-9L &	10  & 160,000		& 109 (0.6MB) 	& 0.068 &  84.5\% \\ \hline
         GTSRB-RN34 	&	43  & 	100,000	& 47	 (1.7MB) & 0.047 &  98.1\% \\
         CIFAR10-RN34 &	10  & 100,000		& 109 (3.9MB) &	0.109 &  94.6\% \\
         Caltech-RN34 &	256  & 100,000		& 27	 (1.0MB) &	0.027 &  74.4\% \\ \hline
    \end{tabular}}
\end{center}
\end{table}

\noindent\textbf{Utility for legitimate clients:} Suppose we want a confidence for reliable demonstration of ownership equal to $1- 2^{-64}$.
$F_\mathcal{V}$ has a prediction API with 1M \client{s} (1M watermarks are registered for $F_\mathcal{V}$). We need $P(L < e) < 10^{-6} \times 2^{-64} = 5.4 \times 10^{-26}$ to be able to test all registered watermarks while achieving our targeted confidence.
We choose a tolerated error rate $e=0.5$.
Table~\ref{tab:WM_size} reports the computed watermark ratio $r_w$ required to protect six models against model extraction.
We see $r_w$ must always be lower than 0.5\% to reach $1- 2^{-64}$ confidence for any victim model.
$F_\mathcal{V}$'s accuracy is thus degraded in a negligible manner that does not impact its utility. \ourname meets the reliability~\ref{req:non-trivial} and utility~\ref{req:qos} requirements.

\noindent\textbf{Overhead:} Storing 1M watermarks would require at most a few TBs (cf. Tab.~\ref{tab:WM_size}). Watermark verification consists in obtaining predictions from a purported surrogate model. It is operated by \judge who gets predictions at no monetary cost.
Thus, demonstration of ownership is only a matter of time and getting one prediction from our most complex model (Caltech-RN34) takes 9ms (on Tesla P100 GPU).
Verifying one watermark for this model takes 0.25s (27 queries) and verifying 100,000 watermarks takes 7 hours using a single GPU.
\judge can initially verify all watermarks with a lower confidence to reduce this time (by testing only a subset of each watermark). Only successful verification would later undergo a verification of the full watermark.
Testing the same 100,000 watermarks with $1- 2^{-16}$ targeted confidence (instead of $1- 2^{-64}$) requires 1h15 (5 samples per watermark).
This time can further be reduced by parallelizing predictions on several GPUs.
\ourname's verification process is more computationally expensive due to the requirement of testing all watermarks to account for Sybils. However, unlike prior watermarking schemes, \ourname is effective against model extraction attacks.

\subsection{Effectiveness against real extraction attacks}
\label{sec:real-attack}

\begin{table}
\begin{center}
    \caption{Efficacy of \ourname to defend against PRADA and KnockOff model extraction attacks. Baseline gives the test accuracy $Acc_{test}$ of the victim $F_\mathcal{V}$ and surrogate model $F_\mathcal{A}$ trained without \ourname in place. $F_\mathcal{A}$ with \ourname provides $Acc_{test}$ and watermark accuracy $Acc_{wm}$ of $F_\mathcal{A}$ when \ourname protects $F_\mathcal{V}$. All $F_\mathcal{A}$ have high $Acc_{wm} > 0.5$ allowing successful demonstration of ownership.}
    \label{tab:stealing-complex}
    \resizebox{1.\columnwidth}!{
    \begin{tabular}{llcccc} \hline
                    &	            & \multicolumn{2}{c}{Baseline $Acc_{test}$} & \multicolumn{2}{c}{$F_\mathcal{A}$ with \ourname} \\
        Attack      & Model	        & $F_\mathcal{V}$   & $F_\mathcal{A}$	    & $Acc_{test}$  & $Acc_{wm}$  \\ \hline
                    & MNIST-5L      & 98.71\%           & 95\%                  & 78.93\%       & 100.00\% \\
        PRADA       & GTSRB-5L      & 91.50\%           & 61.00\%               & 61.43\%       & 98.23\% \\
                    & CIFAR10-9L    & 84.53\%           & 60.03\%               & 60.95\%       & 71.17\% \\ \hline
                    & GTSRB-RN34    & 98.42\%           & 97.43\%               & 97.72\%       & 100.00\% \\
        KnockOff    & CIFAR10-RN34  & 94.66\%           & 88.27\%               & 88.41\%       & 72.54\% \\
                    & Caltech-RN34  & 74.62\%           & 72.74\%               & 71.98\%       & 93.54\% \\ \hline
    \end{tabular}}
\end{center}
\end{table}

We want to show that any surrogate $F_\mathcal{A}$ of a victim model $F_\mathcal{V}$ protected by \ourname will embed a watermark that allows for reliable demonstration of ownership.
We evaluate the effectiveness of \ourname against two landmark model extraction attacks namely PRADA~\cite{juuti2019prada} and KnockOff~\cite{orekondy2018knockoff}.

Low-capacity models expose a prediction API that returns prediction classes $\hat{F}_\mathcal{V}$ required for the PRADA attack. High-capacity models return the full probability vector $F_\mathcal{V}$.
Each victim model is protected by \ourname using the setting presented in Sect.~\ref{sec:dawn_settings}.
This setting enables \victim to demonstrate ownership of each surrogate model with confidence $1-2^{-64}$ using a tolerated error rate $e=0.5$.
For demonstration of ownership to be successful, the surrogate model $F_\mathcal{A}$ must pass the watermark verification test $L(T_\mathcal{A},\hat{B}_\mathcal{V}(T_\mathcal{A}),F_\mathcal{A}) < e$.
In our setting, it means that \ourname successfully defends against an extraction attack if the watermark accuracy for $F_\mathcal{A}$ is larger than 50\%, i.e., $Acc_{wm}(F_\mathcal{A}) > 1-e$.

Table~\ref{tab:stealing-complex} presents the result of this experiment. We see all surrogate models have a watermark accuracy $Acc_{wm} \geq 50\%$, which means \victim is successful in demonstrating their ownership.
\ourname successfully defends against the PRADA and KnockOff attacks for all tested models while incurring little decrease in $F_\mathcal{V}$'s utility (evaluated in Sect.~\ref{sec:dawn_settings}).
We have shown \ourname effectively embeds a watermark in surrogate models $F_\mathcal{A}$ stolen using extraction attacks.
In Table~\ref{tab:stealing-complex}, note that \ourname significantly decreases the surrogate model test accuracy ($Acc_{test}$ for $F_{\mathcal{A}}$) for MNIST-5L while it has little impact on the same for other datasets.
Drastic reduction in $Acc_{test}$ is not a concern from the defender's perspective - in fact it can, by itself, serve as a deterrence for \adv against model extraction. In all cases, adequate watermark accuracy $Acc_{wm} > 0.5$ serves as a deterrence.

\subsection{Resilience to distributed extraction attack}
\label{sec:resilience-distributed}
Distributing a model extraction attack across several \client{s} means several adversaries $\mathcal{A}_i$ query a subset $D_{\mathcal{A}_i}$ from the whole set $D_{\mathcal{A}}$ used to train the surrogate model $F_\mathcal{A}$.
Recall that \ourname is a deterministic mechanism Sect.~\ref{sec:wm-generation}.
The watermarking $W_\mathcal{V}$ and backdoor $B_\mathcal{V}$ functions are deterministic and specific to $F_\mathcal{V}$.
Their results only depend on the input queried to $F_\mathcal{V}$.
The responses to $D_\mathcal{A}$, and its corresponding trigger set, remain the same regardless of which client(s) query the prediction API.
Thus, $D_\mathcal{A}$ is labeled in the same manner and it includes the same trigger set $T_\mathcal{A}$ whether it is queried by one or by multiple \client{s}.
Thus, the watermark in $F_\mathcal{A}$ trained using $D_\mathcal{A}$ will remain indistinguishable~\ref{req:indifferentiability} and unremovable~\ref{req:unremovability} even if multiple clients collude.

Note that in the case of colluding clients, each adversary $\mathcal{A}_i$ has a subset $T_{\mathcal{A}_i}$ of the whole trigger set $T_\mathcal{A}$.
When verifying ownership, the judge \judge will have several successful watermark verifications
$L(T_{\mathcal{A}_i},B_\mathcal{V}(T_{\mathcal{A}_i}),F_\mathcal{A}) < e$: one for each adversary $\mathcal{A}_i$ who colluded to build the surrogate model $F_\mathcal{A}$.
The verification of each sub-watermark $(T_{\mathcal{A}_i},B_\mathcal{V}(T_{\mathcal{A}_i}))$ has the same expectation for success as the verification of the whole watermark $(T_\mathcal{A},B_\mathcal{V}(T_\mathcal{A}))$.
\judge will conclude that each \client $i$ whose watermark is successfully verified is a perpetrator of the distributed extraction attack used to build the surrogate model $F_\mathcal{A}$.
Linkability~\ref{req:linkability} remains valid in case of collusion.

In a distributed attack, the watermark associated to each colluding client is smaller than in a centralized attack. To verify ownership with a same reliability, we must increase the watermark size and consequently $r_w$ by a factor equal to the number of colluding clients.
We assume the number of real colluding clients is limited, e.g., a few tens. Nevertheless, it is possible to mount a Sybil attack in which several API accounts are created by a single adversary.
The API account registration process must require providing information that maximizes difficulty of creating trusted accounts, e.g., verified phone number or credit card, to mitigate this threat.
Also, Sybils-detection techniques exist~\cite{tran2009boundedsybils,wang2013sybils-clustering} and it is possible to link Sybils accounts by examining querying patterns and IP addresses for instance~\cite{stringhini2015sybils-clustering-act}.
For example, to protect Caltech-RN34, we could increase $r_w$ to reliably verify the watermark of 35 colluders while maintaining the utility loss below 1\%.
Consequently, the higher the number of classes, the greater the reliability of watermark verification (c.f. Eq.~\ref{eq:proof}) and we can tolerate more Sybils. For a classifier with 10,000 classes and utility loss below 1\% we can reliably verify the watermark of 87 colluders.

\section{Watermark Removal}
\label{sec:watermark-removal}

Several techniques can identify if a DNN model has a backdoor~\cite{chen2018detecting,guo2019tabor,wang2019neural}.
Most techniques like \textit{Neural Cleanse}~\cite{wang2019neural} and \textit{TABOR}~\cite{guo2019tabor} can only detect backdoors for which the trigger is a static pattern added to original inputs (e.g., yellow square added to an image).
In contrast, our trigger set is composed of unmodified samples having only incorrect labels.
Consequently, techniques like \textit{Neural Cleanse} and \textit{TABOR} are ineffective at detecting \ourname watermark.
In this section, we evaluate the resilience of \ourname watermarks to removal using six attacks: (1) double-extraction of a \emph{second order surrogate model} $F'_\mathcal{A}$, (2) fine-tuning~\cite{kornblith2018better}, (3) pruning~\cite{blalock2020pruning}, (4) training with noise, (5) adding noise during the inference, and (6) recognizing queries from training data.

\textbf{Double extraction and fine-tuning.}\label{sec:double-stealing}
A watermark may be removed by performing an extraction attack against $F_\mathcal{A}$ to obtain a second order surrogate model $F_\mathcal{A}'$.
\adv has full control over $F_\mathcal{A}$: its prediction API is not protected by \ourname and does not intentionally return incorrect prediction.
If \adv uses a disjoint set of queries to extract a surrogate $F_\mathcal{A}'$ from $F_\mathcal{A}$, then $F_\mathcal{A}'$ may not embed the watermark, preventing the demonstration of its ownership by \victim.
Instead of starting the second extraction from scratch, it can use $F_\mathcal{A}$ as the starting point for $F_\mathcal{A'}$ and fine-tune it.
We call this \emph{stealing+fine-tuning}.

We observed in Tab.~\ref{tab:stealing-complex} that surrogate models have a lower accuracy than victim models because model extraction incurs a necessary decrease in surrogate model accuracy.
We evaluate the extent of the decrease in $Acc_{test}$ and  $Acc_{wm}$ if \adv launches two successive extraction attacks instead of one or steals+fine-tunes the model to obtain $F_\mathcal{A}'$: the first against $F_\mathcal{V}$ and the second against $F_\mathcal{A}$.
We evaluate these evasion techniques using the PRADA and KnockOff attacks.

The two successive extraction attacks and stealing+fine-tuning are performed in the same conditions as the first attack.
The only difference is that \adv uses half the seed samples for each PRADA attack (5 per class for MNIST-5L and  GTSRB-5L, 500 per class for CIFAR10-9L) and runs an additional duplication round to query the same number of inputs.
The number of seed samples is a limited adversarial capability in PRADA~\cite{juuti2019prada}, so we grant \adv with the same capability for single and double extraction attack.
For each KnockOff attack, \adv uses a different set of 100,000 inputs from ImageNet.
For the second extraction attack (against $F_\mathcal{A}$), we query the same number of inputs as for the first one. While this number can be increased, we empirically observed that the test accuracy of $F_\mathcal{A}'$ reaches its maximum and stagnates before the PRADA and KnockOff attacks finish, i.e., more queries do not improve $Acc_{test}$ of $F_\mathcal{A}'$.

\begin{table}[htb]
    \begin{center}
        \caption{Double extraction attack: test accuracy $Acc_{test}$ for the victim model $F_\mathcal{V}$, first order surrogate model $F_\mathcal{A}$ and second order surrogate model $F_\mathcal{A}'$. Watermark accuracy $Acc_{wm}$ for $F_\mathcal{A}'$ and increase in $Acc_{test}$ degradation (Degrad.) between $F_\mathcal{A}$ and $F_\mathcal{A}'$ compared to between $F_\mathcal{V}$ and $F_\mathcal{A}$. Double extraction can remove \ourname watermarks from $F_\mathcal{A}'$ but it incurs a significant additional degradation in test accuracy (20-80\%). Test accuracy of $F_\mathcal{A}'$ for the PRADA attack is too low for it to be useful.}
        \label{tab:double-stealing}
        \resizebox{1.\columnwidth}!{
        \begin{tabular}{lccccc} \hline
                            & $F_\mathcal{V}$		& $F_\mathcal{A}$	& \multicolumn{3}{c}{$F_\mathcal{A}'$ (2$^{nd}$ extraction)}	\\
            Model			& $Acc_{test}$ 			& $Acc_{test}$								& $Acc_{test}$		& $Acc_{wm}$								& Degrad. \\ \hline
            MNIST-5L		& 98.7\%				& 71.10\%			& 49.06\% (-49pp)	& 11.11\% 									& +80\% \\
            GTSRB-5L		& 91.5\%				& 50.66\%				& 29.51\% (-62pp)	& 3.36\% 									& +52\%\\
              CIFAR10-9L		& 84.5\%				& 45.7\%			& 37.80\% (-44pp) 	& 3.04\% 									& +20\% \\ \hline
              GTSRB-RN34 		& 98.42\%     			& 97.72\%			& 97.30\% (-1pp)	& 17.64\%	&							+71\% \\
              CIFAR10-RN34 	& 94.66\% 				& 88.41\% 				& 84.85\% (-10pp)	& 11.88\%		&					+57\% \\
              Caltech-RN34 	& 74.62\% 				& 71.98\% 			& 70.95\% (-4pp) 	& 9.67\%									& +39\%
        \end{tabular}}
    \end{center}
    \end{table}

    \begin{table}[htb]
    \begin{center}
        \caption{Stealing+fine-tuning attack: test accuracy $Acc_{test}$ for the victim model $F_\mathcal{V}$, surrogate model $F_\mathcal{A}$ and fine-tuned surrogate model $F_\mathcal{A}'$. Watermark accuracy $Acc_{wm}$ for $F_\mathcal{A}'$ and increase in $Acc_{test}$ degradation (Degrad.) between $F_\mathcal{A}$ and $F_\mathcal{A}'$ compared to between $F_\mathcal{V}$ and $F_\mathcal{A}$. Fine-tuning can remove \ourname watermarks from $F_\mathcal{A}'$ but it incurs additional degradation in test accuracy (20-60\%). Test accuracy of $F_\mathcal{A}'$ for the PRADA attack is too low for it to be useful.}
        \label{tab:fine-tuning}
        \resizebox{1.\columnwidth}!{
        \begin{tabular}{lccccc} \hline
                            & $F_\mathcal{V}$		& $F_\mathcal{A}$	& \multicolumn{3}{c}{$F_\mathcal{A}'$ (fine-tuning)}	\\
            Model			& $Acc_{test}$ 			& $Acc_{test}$								& $Acc_{test}$		& $Acc_{wm}$								& Degrad. \\ \hline
            MNIST-5L		& 98.7\%				& 71.10\%			& 63.67\% (-35pp)	& 22.22\%									& +27\% \\
            GTSRB-5L		& 91.5\%				& 50.66\%				& 42.09\% (-49pp)	& 9.24\% 									& +21\%\\
              CIFAR10-9L		& 84.5\%				& 45.7\%			& 38.41\% (-46pp)	& 4.06\%									& +19\% \\ \hline
              GTSRB-RN34 		& 98.42\%     			& 97.72\%			& 97.22\% (-1pp)	& 7.81\% 	&							+60\% \\
              CIFAR10-RN34 	& 94.66\% 				& 88.41\% 				& 84.93\% (-10pp)	& 16.83\%		&					+56\% \\
              Caltech-RN34 	& 74.62\% 				& 71.98\% 			& 70.81\% (-4pp)	& 22.58\%							& +44\%
        \end{tabular}}
    \end{center}
    \end{table}

As it can be observed in Tab.~\ref{tab:double-stealing} and~\ref{tab:fine-tuning}, double extraction attack and stealing+fine-tuning effectively remove the watermark from the second order surrogate model $F_\mathcal{A}'$. The watermark accuracy is low enough (3-22\%) to fail demonstration of ownership for $F_\mathcal{A}'$, which empirically confirms that prior DNN watermarking techniques are not resilient to this class of model extraction attacks~\cite{zhang2018protecting}.

While removing the watermark, the extraction of the second order surrogate model using these attacks also increases the degradation in test accuracy by 20\% to 80\% for $F_\mathcal{A}'$  compared to $F_\mathcal{A}$.
Considering a powerful \adv having unlimited access to natural data (e.g., KnockOff adversary model) the extraction of the first order surrogate model incurs little accuracy degradation and so does the extraction of the second order surrogate model. The final model $F_\mathcal{A}'$ stolen using KnockOff attack has its watermark removed and preserves its utility (from -1pp to -10pp compared to $F_\mathcal{V}$). \ourname cannot protect model extraction attacks where \adv has unlimited access to natural data.
However, \adv's access to data is limited in many scenarios, e.g., access to medical imaging that are privacy sensitive, and highly specialized models may not return meaningful predictions to random images. In such a scenario, the KnockOff attack may not be effective and the PRADA attack is more effective. We see that for the PRADA attack the test accuracy of $F_\mathcal{A}'$ decreases sharply during each attack that removes the watermark (3 top rows in
Tab.~\ref{tab:double-stealing} and~\ref{tab:fine-tuning}) .
In most cases, the final test accuracy of $F_\mathcal{A}'$ is less than half of $F_\mathcal{V}$ (from -35pp to -62pp) and we consider that $Acc(F_\mathcal{A}') \ll Acc(F_\mathcal{V})$ makes $F_\mathcal{A}'$ too inaccurate to be useful.
\ourname can effectively protect a model against extraction attack that uses a limited amount of data - it destroys the utility of the model $F_\mathcal{A}'$ deprived of the watermark.

Double-extraction and fine-tuning can remove the watermark while preserving test accuracy given that \adv has unlimited access to natural data.
However, \ourname defeats both attacks when \adv has limited access to data, which is the case for most model extraction attack scenarios (cf. Sect.~\ref{sec:bg_extraction}).
\adv's access to data is limited in many scenarios, e.g., medical imaging classifiers, where these removal attacks would be ineffective.

\textbf{Pruning.} Alternatively, \adv may attempt to prune the model by setting random weights of the model to zero.
In our experiments we prune weights uniformly randomly.
We show that for large values of $\delta$ pruning is effective at removing the watermark but it sacrifices model's utility and renders it useless (c.f. Table~\ref{tab:pruning}).
Furthermore, we observe that $Acc_{test}$ and $Acc_{wm}$ do not fall proportionally i.e. there is no guarantee that sacrificing X\% $Acc_{test}$ results in the same drop in $Acc_{wm}$.
Also, our experiments show that determining appropriate $\delta$ without knowing $T_\mathcal{A}$ is challenging as there is no consistent drop in accuracy for a particular value of $\delta$ across all models.

\setlength\tabcolsep{4pt}
\begin{table}
\begin{center}
    \caption{$Acc_{test}$ and $Acc_{wm}$ of pruned stolen models $F_\mathcal{A}$. $\delta$ denotes the percentage of weights set to zero. $Acc_{wm} > 50\%$ ensures successful demonstration of ownership. We consider a decrease in $Acc_{test} > 10pp$ an unacceptable loss of utility of $F_{\mathcal{A}}$. Purple (underline) results highlight low $Acc_{wm}$ and low $Acc_{test}$: $F_\mathcal{A}$ is unusable. Red (dashed underline) results highlight low $Acc_{wm}$ while $Acc_{test}$ remains high: \victim may fail to prove ownership.}
    \label{tab:pruning}
    \resizebox{1.\columnwidth}!{
    \begin{tabular}{ll|cccccc} \hline
            \multicolumn{2}{c|}{}           & \multicolumn{6}{c}{$\delta$ (\%)}\\
            Model           &  				& 0         & 25        & 40        & 50        & 75        & 90\\ \hline
            MNIST-5L		& $Acc_{test}$	& 78.93\%   & 78.57\%   & 78.31\%   & 77.76\%   & 71.24\%   & \accallfail{43.84\%}\\
                            & $Acc_{wm}$	& 100.00\%  & 100.00\%  & 100.00\%  & 100.00\%  & 61.50\%   & \accallfail{7.69\%}\\ \hline
            GTSRB-5L     	& $Acc_{test}$	& 61.43\%   & 54.24\%   & 43.80\%   & \accallfail{30.57\%}   & \accallfail{3.81\%}    & \accallfail{3.80\%}\\
                            & $Acc_{wm}$	& 98.23\%   & 90.26\%   & 50.44\%   & \accallfail{20.35\%}   & \accallfail{3.53\%}	  & \accallfail{3.53\%}\\ \hline
            CIFAR10-9L      & $Acc_{test}$	& 60.95\%   & 59.83\%   & 48.45\%   & \accallfail{44.96\%}   & \accallfail{10.00\%}   & \accallfail{10.00\%}\\
                            & $Acc_{wm}$	& 71.17\%   & 70.20\%   & 54.91\%   & \accallfail{16.58\%}   & \accallfail{16.26\%}   & \accallfail{11.43\%}\\ \hline
            GTSRB-RN34      & $Acc_{test}$	& 97.72\%   & 97.72\%   & 97.71\%   & \accwmfail{97.41\%}   & \accallfail{45.73\%}   & \accallfail{0.47\%}\\
                            & $Acc_{wm}$	& 72.54\%   & 62.74\%   & 50.98\%   & \accwmfail{33.34\%}   & \accallfail{11.62\%}   & \accallfail{1.96\%}\\ \hline
            CIFAR10-RN34    & $Acc_{test}$	& 88.41\%   & 87.85\%   & 87.12\%   & 86.48\%   & \accallfail{36.69\%}   & \accallfail{10.00\%}\\
                            & $Acc_{wm}$	& 100.00\%  & 99.00\%   & 91.08\%   & 81.18\%   & \accallfail{7.92\%}    & \accallfail{4.95\%}\\ \hline
            Caltech-RN34	& $Acc_{test}$  & 71.98\%   & 71.06\%   & 68.48\%   & 63.55\%   & \accallfail{4.92\%}    & \accallfail{0.33\%}\\
                            & $Acc_{wm}$	& 93.54\%   & 93.54\%	& 70.96\%   & 61.29\%   & \accallfail{6.45\%}    & \accallfail{0.00\%}\\
    \end{tabular}}
\end{center}
\end{table}

\textbf{Training with noise.}
\adv may attempt to weaken the embedding of the watermark by adding some noise to the samples before they start training.
\adv's goal is not to expose the model to the samples that would be eventually used for the verification.
We show that for the values up to $\epsilon = 0.4$, $Acc_{wm}$ is not affected (c.f. Table~\ref{tab:noisytraining}).
Beyond that in all but one case, either $Acc_{wm}$ remains above the effectiveness threshold, or the drop in $Acc_{test}$ becomes unacceptable.
This approach is thus not effective at successfully stealing the model while circumventing \ourname.

\setlength\tabcolsep{4pt}
\begin{table}
\begin{center}
    \caption{$Acc_{test}$ and $Acc_{wm}$ of stolen models $F_\mathcal{A}$ when trained with $\epsilon$-perturbed samples. $Acc_{wm} > 50\%$ ensures successful demonstration of ownership. We consider a decrease in $Acc_{test} > 10pp$ an unacceptable loss of utility of $F_{\mathcal{A}}$. Purple (underline) results highlight low $Acc_{wm}$ and low $Acc_{test}$: $F_\mathcal{A}$ is unusable. Red (dashed underline) results highlight low $Acc_{wm}$ while $Acc_{test}$ remains high: \victim may fail to prove ownership.}
    \label{tab:noisytraining}
    \resizebox{1.\columnwidth}!{
    \begin{tabular}{ll|ccccc} \hline
            \multicolumn{2}{c|}{}           & \multicolumn{5}{c}{$\epsilon$ perturbation}\\
            Model           &  				& $F_\mathcal{V}$ & 0.0         & 0.25     & 0.4        & 0.75      \\ \hline 
            MNIST-5L		& $Acc_{test}$	& 98.71\%         & 78.93\%     & 76.44\%  & 79.99\%    & 72.00\%   \\ 
                            & $Acc_{wm}$	& X               & 100.00\%    & 100.00\% & 100.00\%   & 100.00\%  \\ \hline
            GTSRB-5L     	& $Acc_{test}$	& 91.50\%         & 61.43\%     & 54.85\%  & 49.90\%    & \accallfail{29.60\%}   \\ 
                            & $Acc_{wm}$	& X               & 98.23\%     & 95.24\%  & 78.46\%    & \accallfail{25.53\%}   \\ \hline
            CIFAR10-9L      & $Acc_{test}$	& 84.53\%         & 60.95\%     & 59.39\%  & 54.84\%    & \accallfail{44.59\%}   \\ 
                            & $Acc_{wm}$	& X               & 71.17\%     & 71.39\%  & 67.70\%    & \accallfail{29.33\%}   \\ \hline
            GTSRB-RN34      & $Acc_{test}$	& 98.42\%         & 97.72\%     & 97.41\%     & 97.01\%       & 97.07\%      \\ 
                            & $Acc_{wm}$	& X               & 72.54\%     & 66.67\%     & 58.82\%       & 54.90\%      \\ \hline
            CIFAR10-RN34    & $Acc_{test}$	& 94.66\%         & 88.41\%     & 87.87\%     & 86.90\%       & 85.76\%      \\ 
                            & $Acc_{wm}$	& X               & 100.00\%    & 100.00\%     & 91.08\%       & 73.26\%      \\ \hline
            Caltech-RN34	& $Acc_{test}$  & 74.62\%         & 71.98\%     & 70.71\%     & 69.68\%       & \accwmfail{67.25\%}      \\ 
                            & $Acc_{wm}$	& X               & 93.54\%     & 74.19\%     & 74.19\%       & \accwmfail{48.38\%}      \\ 
    \end{tabular}}
\end{center}
\end{table}

\textbf{Inference with noise.}
Instead of training with noisy samples, \adv can add noise to all samples during the inference in attempt to avoid verification. However, this will reduce utility for \adv's clients. In Tab.~\ref{tab:noisy-verification}, we show the decrease in $Acc_{wm}$ and corresponding $Acc_{test}$ for various amounts of noise $\epsilon$. We show that in almost all cases $Acc_{wm}$ remains high or $Acc_{test}$ drops below the acceptable utility level (purple, underline). In few cases (red, dashed underline) $Acc_{test}$ remains high while $Acc_{wm}$ decreases to $<50\%$. However, the $\epsilon$ value that benefits \adv the most is not consistent across the models. Hence, for a particular value of $\epsilon$ that preserves $Acc_{test}$, \adv has no guarantee that watermark verification would fail.

\textbf{Recognizing queries from training data.}
Alternatively, \adv having deployed $F_\mathcal{A}$ and being aware of \ourname can try to prevent watermark verification performed by \judge. \adv can check if queries to $F_\mathcal{A}$ belong to $D_\mathcal{A}$ used to steal $F_\mathcal{V}$, and return different predictions for them.
We evaluated that searching $D_\mathcal{A}$ for exact matches can be done efficiently using a hash table (28-44ms additional overhead per query for our datasets).
Hashing the query is the most time consuming part of this search.

To prevent this evasion by \adv, \judge can slightly perturb samples in $T_\mathcal{A}$ before submitting them for verification to $F_\mathcal{A}$.
In Tab.~\ref{tab:noisy-verification}, we show the resilience of watermark verification to various amounts of noise ($\epsilon$) added to each image in $T_\mathcal{A}$.
We show that we can maintain $Acc_{wm} > 50\%$ for all models up to $\epsilon = 0.25$ or even more for some models.
\adv can no longer perform a simple lookup in a hash table to identify queries from \judge  if $T_\mathcal{A}$ is perturbed. \adv must compute the distance to every sample in $D_\mathcal{A}$ and find the nearest neighbor to the query.
We demonstrate that such search incurs a substantial computational overhead (c.f. Tab.~\ref{tab:timings}). Searching in 100,000 ImageNet samples can take over 7s per query (on a server-grade machine with a Xeon CPU and 64 GB of RAM), which is too long to be acceptable for deployment.
Also, a small distance to an element in $D_\mathcal{A}$ does not mean that queried image was in fact part of $D_\mathcal{A}$. \adv has to set up a threshold for rejecting the queries that will affect the utility of the model, on top of the described computational overhead.

\setlength\tabcolsep{4pt}
\begin{table}
\begin{center}
    \caption{$Acc_{wm}$ and $Acc_{test}$ when \adv with added $\epsilon$ perturbation during inference. $Acc_{wm} > 50\%$ ensures successful demonstration of ownership. We consider a decrease in $Acc_{test} > 10pp$ an unacceptable loss of utility of $F_{\mathcal{A}}$. Purple (underline) results highlight low $Acc_{wm}$ and low $Acc_{test}$: $F_\mathcal{A}$ is unusable. Red (dashed underline) results highlight low $Acc_{wm}$ while $Acc_{test}$ remains high: \victim may fail to prove ownership.}
    \label{tab:noisy-verification}
    \resizebox{1.\columnwidth}!{
    \begin{tabular}{ll|ccccccccc} \hline
            \multicolumn{2}{c|}{}               & \multicolumn{7}{c}{$\epsilon$ perturbation}\\
            Model               &               & 0.0       & 0.1       & 0.25      & 0.4       & 0.5       & 0.75      & 0.9\\ \hline
            MNIST-5L            & $Acc_{test}$  & 78.93\%   & 78.66\%   & 78.62\%   & 78.42\%   & 77.45\%   & 76.89\%   & \accallfail{66.21\%}\\
                                & $Acc_{wm}$	& 100\%     & 100\%     & 100\%     & 100\%     & 92.30\%   & 53.84\%   & \accallfail{30.76\%}\\
            GTSRB-5L            & $Acc_{test}$  & 61.43\%   & 60.72\%   & 42.36\%   & 37.71\%   & \accallfail{34.56\%}   & \accallfail{27.27\%}   & \accallfail{23.29\%}\\
                                & $Acc_{wm}$	& 98.23\%   & 96.46\%   & 88.49\%   & 66.37\%   & \accallfail{37.16\%}   & \accallfail{10.61\%}   & \accallfail{4.42\%} \\
            CIFAR10-9L          & $Acc_{test}$  & 60.95\%   & 60.19\%   & 52.89\%   & \accallfail{40.4\%}    & \accallfail{31.51\%}   & \accallfail{18.43\%}   & \accallfail{14.97\%}\\
                                & $Acc_{wm}$	& 71.17\%   & 69.56\%   & 51.36\%   & \accallfail{31.56\%}   & \accallfail{19.80\%}   & \accallfail{7.89\%}    & \accallfail{7.56\%} \\
            GTSRB-RN34          & $Acc_{test}$  & 97.72\%   & 97.63\%   & 96.51\%   & \accwmfail{93.97\%}   & \accwmfail{90.91\%}   & \accallfail{84.01\%}   & \accallfail{79.04\%}\\
                                & $Acc_{wm}$	& 72.54\%   & 66.67\%   & 54.90\%   & \accwmfail{29.41\%}   & \accwmfail{21.56\%}   & \accallfail{11.76\%}   & \accallfail{9.80\%} \\
            CIFAR10-RN34        & $Acc_{test}$  & 88.41\%   & 88.21\%   & 86.35\%   & 82.79\%   & \accwmfail{79.84\%}   & \accallfail{66.96\%}   & \accallfail{56.8\%}\\
                                & $Acc_{wm}$	& 100\%     & 99.00\%   & 92.07\%   & 67.32\%   & \accwmfail{39.60\%}   & \accallfail{19.08\%}   & \accallfail{8.91\%} \\
            Caltech-RN34        & $Acc_{test}$  & 71.98\%   & 71.45\%   & 65.52\%   & \accallfail{54.57\%}   & \accallfail{45.14\%}   & \accallfail{23.07\%}   & \accallfail{14.45\%}\\
                                & $Acc_{wm}$	& 93.54\%   & 90.32\%   & 51.61\%   & \accallfail{19.35\%}   & \accallfail{6.45\%}    & \accallfail{0.00\%}    & \accallfail{0.00\%}
    \end{tabular}}
\end{center}
\end{table}

\setlength\tabcolsep{4pt}
\begin{table}
\begin{center}
    \caption{Time to find the closest sample in the dataset (ms), average over 10,000 tests (negligible standard deviation). Exact lookup (Eq.) using a hash table is done in negligible time ($\approx 30ms$). Calculating distances has a substantial time overhead.}
    \label{tab:timings}
    \begin{tabular}{l|cccccc} \hline
        Dataset 	& samples	& size	    & Eq.& $L_0$ & $L_2$ & $L_{\inf}$ \\ \hline
        MNIST  		& 60,000    & 28x28x1	& 28 & 230   & 770   & 570 \\
        GTSRB		& 39,209   	& 32x32x3   & 29 & 172   & 420   & 420 \\
		CIFAR10		& 50,000  	& 32x32x3   & 30 & 243   & 625   & 615 \\
		Caltech	 	& 23,703	& 224x224x3 & 30 & 351   & 2020  & 2040 \\
		ImageNet	& 100,000   & 224x224x3 & 44 & 1421  & 7362  & 7370
    \end{tabular}
\end{center}
\end{table}

\section{Discussion}
\label{sec:discussion}

\subsection{Meeting system requirements}
\noindent\textbf{Unremovability~\ref{req:unremovability}.}
We extensively evaluated (Sect~\ref{sec:unremovability}) that manipulation of the training process of $F_\mathcal{A}$ either does not prevent the embedding of the watermark or if it does, it significantly degrades $F_\mathcal{A}$'s test accuracy. Proper use of regularization can effectively mitigate the  watermark embedding but it requires \adv to be granted more capabilities e.g. increased access to relevant data.
We also showed (Sect.~\ref{sec:double-stealing}) that two successive extraction attacks can remove a watermark from $F_\mathcal{A}$. However, it also decreases the test accuracy to an extent that makes $F_\mathcal{A}$ unusable.
Finally, prior work~\cite{adi2018turning,merrer2017adversarial,zhang2018protecting} has shown that manipulations after training such as pruning and adversarial fine-tuning are ineffective against backdoor-based DNN watermarks.
Our evaluation confirmed that watermark cannot be removed using pruning or fine-tuning (Sect.~\ref{sec:watermark-removal}) without sacrificing the utility of the model.
We can conclude that \ourname watermarking meets unremovability requirement.

\noindent\textbf{Indistinguishability~\ref{req:indifferentiability}.}
We defined model-specific watermarking and backdoor functions (Sect. \ref{sec:wm-generation}) that \emph{always} return the same same output (correct or incorrect) for the same input.
We also introduced a solution for mapping inputs with minor differences to similar predictions (Sect~\ref{sec:indistinguishability}).

\noindent\textbf{Reliability~\ref{req:non-trivial} and utility~\ref{req:qos}.}
The watermark registration and verification protocol that we introduce (Sect.~\ref{sec:registration-verification}) ensures that the success of \adv in demonstrating ownership of an arbitrary model is negligible.
We showed how to set up \ourname in order to reliably demonstrate ownership of several surrogate models stolen using two state-of-the-art model extraction attacks with high confidence equal to $1- 2^{-64}$ (Sect.~\ref{sec:eval-stealing}).
\ourname effectively watermarked every surrogate model $F_\mathcal{A}$ while causing a negligible decrease of $F_\mathcal{V}$'s utility (0.03-0.5\%).
\ourname allows for reliable ownership demonstration while preserving the utility.

\noindent\textbf{Non-ownership piracy~\ref{req:non-piracy}} is guaranteed by our watermark registration and verification protocol (Sect.~\ref{sec:registration-verification}). In case of contention, the first registered model is deemed the original.

\noindent\textbf{Linkablity~\ref{req:linkability}.}
\ourname selects watermarked inputs from \client queries and registers one watermark per \client. Different clients make different queries and they will consequently have different watermarks.
Given that we meet the reliability requirement~\ref{req:non-trivial}, a single watermark $(T_{\mathcal{A}_i},B_\mathcal{V}(T_{\mathcal{A}_i}))$ will succeed in proving $F_{\mathcal{A}_i}$ is a surrogate of $F_\mathcal{V}$.
Watermarks are \client-specific which makes a surrogate model linkable to an \client.

\noindent\textbf{Collusion resistance~\ref{req:sybils}.}
\ourname relies on deterministic functions for watermarking ($W_\mathcal{V}$) and backdooring ($B_\mathcal{V}$) that are specific to $F_\mathcal{V}$ but independent of the client sending a query.
Thus, the watermark remains indistinguishable despite collusion~\ref{req:indifferentiability}.
\ourname is resilient to a Sybil attack (bounded to a certain number of Sybils) and assure successful verification by \judge (Sect.~\ref{sec:resilience-distributed}).

\subsection{Limitations}
\adv can attempt to prevent ownership demonstration for $F_\mathcal{A}$ by ensuring that watermark verification (Eq.~\ref{eq:verification}) fails.
This entails reducing the watermark accuracy by training $F_\mathcal{A}$ using only a subset of the trigger set $T_\mathcal{A}$.
Since watermarked inputs are indistinguishable (\ref{req:indifferentiability}) and uniformly distributed in $D_\mathcal{A}$, \adv cannot selectively discard them.
Nevertheless, \adv can discard $x\%$ of the whole $D_\mathcal{A}$, which statistically, will result in $x\%$ of watermarked input being discarded.
This should reduce $Acc_{wm}(F_\mathcal{A})$ by $100-x\%$. If $x$ is high enough, the resulting $Acc_{wm}(F_\mathcal{A})$ can be brought down low enough for watermark verification to systematically fail.

While this strategy is effective, it deprives \adv from a large part of $D_\mathcal{A}$.
This decreases $Acc_{test}(F_\mathcal{A})$ and consequently the utility of $F_\mathcal{A}$ (Sec.~\ref{sec:double-stealing}).
Alternatively, \adv must collect a set $D_\mathcal{A}$ $x\%$ larger and make $x\%$ more queries to $F_\mathcal{V}$ to compensate for later discarded training inputs.
We already discussed in Sect~\ref{sec:bg_extraction} that \emph{access to relevant data is the main limitation} for \adv.
The secondary goal of \adv is to limit the number of queries to $F_\mathcal{V}$ (cf. Sect.~\ref{sec:adv_model}).
This evasion strategy requires more adversarial capabilities (access to data) and it compromises one adversary goal (minimum number of queries). Consequently, even if effective, we do not consider it a realistic evasion strategy.

Another potential limitation of \ourname is circumvention of the mapping function $M_\mathcal{V}$ (Sec.~\ref{sec:indistinguishability}).
If mapping is too aggressive, \adv may probe the input space and try to identify subspaces that are grouped together.
However, this is not guaranteed to work because the behavior of the model on synthetic samples is undefined~\cite{goodfellow2014explaining}.
$M_\mathcal{V}$ impacts only the watermarking decision $W_\mathcal{V}$ and not the returned label - \adv cannot interact directly with the mapping function.
On the other hand, if the tolerated modification $\delta$ is too small, \adv might identify watermarked queries by submitting several samples with minor modifications and taking the majority vote of the label. We evaluated this attack in Sect.~\ref{sec:perfect-mapping} showing that mappings from $M_\mathcal{V}$ are as consistent as predictions from $F_\mathcal{V}$.
Semantic-preserving modifications to image queries (e.g., translation, rotation, change in color intensity, etc,) could be used to improve this attack.
However, by using an embedding from $F_\mathcal{V}$ to implement $M_\mathcal{V}$, both functions should be as resilient to semantic-preserving modifications.

\adv can attempt to weaken the embedding of the watermark by adding a small amount of noise to its training samples before starting the training or to all queries during the inference time (Sect.~\ref{sec:watermark-removal}). Although \adv cannot know the optimal $\epsilon$ value that minimizes accuracy loss while rendering watermark verification ineffective, they can choose a loss budget and incur that loss fully (e.g. 10 pp in our examples) - that implies that in Table~\ref{tab:noisy-verification} \adv would have succeeded in two out of the six cases. How to strengthen \ourname against such an adversary that is ready to incur the maximal allowable accuracy loss is still an open problem.

\section{Related Work}
\label{sec:related-work}

\textbf{Watermarking DNN models.}
The first watermarking technique for DNNs~\cite{uchida2017embedding} explicitly embeds additional information into the weights of a DNN after it is trained. Verifying the watermark requires white-box access to the model in order to analyze the weights. A limitation of this approach is that the watermark can be easily removed by minimally retraining the watermarked model.

Alternative approaches~\cite{merrer2017adversarial,adi2018turning,zhang2018protecting,DarvishRouhani:2019:DEW:3297858.3304051,jia2020entangled} that are more robust have been proposed, where the watermark can be verified in a black-box setting. These are based on backdooring and they allow for watermark extraction using only a prediction API, as discussed in Sect.~\ref{sec:bg_wm}. These approaches use both a carefully selected trigger set and a specific training process chosen by the model owner.
The first proposal for such approach~\cite{merrer2017adversarial} consist in modifying the original model boundary using adversarial retraining~\cite{madry2017towards} in order to make the model unique. The watermark is composed of synthetically generated adversarial samples~\cite{goodfellow2014explaining} that are close to the decision boundary.
The impact of selecting a particular distribution for a watermark has been evaluated in~\cite{zhang2018protecting}. It shows that selecting a trigger set from the same distribution as the training data (albeit with minor synthetic modifications) or from a different distribution, does not affect the accuracy of the model for its primary classification task or on its training time, while the watermark gets perfectly embedded.
Finally, more formal foundations and theoretical guarantees for backdoor-based DNN watermarking have been provided in~\cite{adi2018turning}. This work empirically assesses that the removability of a DNN watermark is highly dependent on the training process of the watermarked model (training from scratch vs. re-training).

In contrast to prior DNN watermarking techniques for black-box verification~\cite{merrer2017adversarial,adi2018turning,zhang2018protecting,DarvishRouhani:2019:DEW:3297858.3304051,jia2020entangled}, \ourname considers a victim who (a) does not control the training of the DNN model and (b) cannot select a trigger set $T$ from the whole input space $\mathbb{R}^n$.
\ourname dynamically embeds a watermark in queries made to a model prediction API. Thus, \ourname defends against model extraction attacks and enables a model owner to identify surrogates of its model.

\textbf{Defenses against model extraction.}
It was suggested that the distribution of queries made during an extraction attack is different from benign queries~\cite{juuti2019prada}. Hence, model extraction can be detected using density estimation methods, namely by assessing the ability for queries to fit a Gaussian distribution or not. However, this technique protects only against attacks using synthetic queries and is not effective against, e.g., the KnockOff attack.
Other detection methods analyse subsequent queries close to the classes' decision boundaries~\cite{quiring2018forgottensib, zheng2019blp} or queries exploring abnormally large region of the input space~\cite{Kesarwani2017model}. Both methods are effective but detect only extraction attacks against decision trees. They are ineffective against complex models like DNNs.
Altering predictions returned to \client{s} can mitigate model extraction attacks. Predictions can be restricted to classes~\cite{tramer2016stealing} or adversarially modified to degrade the performance of the surrogate model~\cite{lee2018defending,orekondy2019predictionpoisoning}. However, some extraction attacks~\cite{juuti2019prada} circumvent such defenses because they remain effective using just prediction classes.

Prior defenses to model extraction are designed to protect only simple models~\cite{quiring2018forgottensib,Kesarwani2017model} or to prevent only specific extraction attacks~\cite{lee2018defending,zheng2019blp}.
It is arguable if a generic defense would ever be effective at detecting/preventing model extraction.
Consequently, with \ourname we take a different approach where we assume a surrogate model can be extracted.
Then we propose a generic defense to identify surrogate DNN models that have been extracted from any victim model using any extraction attack.


%


\bibliographystyle{ACM-Reference-Format}
\bibliography{bibliography}

\appendix

\section{Datasets and Models}
\label{appendix:models}

Table~\ref{tab:datasets} presents the characteristics of the datasets we used in our experiments. These are divided into a training and a testing set.
Images were resized to fit the corresponding model architectures used in prior work.
Table~\ref{table:architectures-prada} presents model architectures used for conducting experiments with low capacity models - the perfect-knowledge attacker in Sect.~\ref{sec:unremovability} and reproduction of the PRADA~\cite{juuti2019prada} attack in Sect.~\ref{sec:real-attack}.

\begin{table}[htb]
\begin{center}
	\caption{Image datasets used to evaluate \ourname. Different sample sizes are input to different models.}
	\label{tab:datasets}
	\begin{tabular}{lcccc} \hline
							& 								&					& \multicolumn{2}{c}{Number of samples} \\
		Dataset		& Sample Size			& Classes	& Train		& Test \\ \hline
		MNIST			& 28x28						& 10			& 60,000	& 10,000 \\
		GTSRB			& 32x32 / 224x224	& 43			& 39,209	& 12,630 \\
		CIFAR10		& 32x32 / 224x224	& 10			& 50,000	& 10,000 \\
		Caltech 	& 224x224					& 256			& 23,703	& 6,904 \\ \hline
		ImageNet	& 224x224					& 1000		& 100,000	& - \\
	\end{tabular}
\end{center}
\end{table}

\begin{table}[!hbt]
	\begin{center}
		\caption{Model architectures of low capacity models.}
		\label{table:architectures-prada}
		\begin{tabular}{lcccc} \hline
			\textbf{Layer}	& \textbf{MNIST-3L}	& \textbf{MNIST-5L} & \textbf{GTSRB-5L}	& \textbf{CIFAR10-9L} \\ \hline
			1		& conv2-32			& conv2-32			& conv2-64			& conv2-32 \\
					& maxpool2d			& maxpool2d 		& maxpool2d			& batchnorm2d \\
					& ReLU				& ReLU				& ReLU				& ReLU \\ \hline
			2		& conv2-64			& conv2-64			& conv2-128			& conv2-64 \\
					& maxpool2d			& maxpool2d			& maxpool2d			& ReLU \\
					& ReLU				& ReLU				& ReLU				& maxpool2d \\ \hline
			3		& dropout			& conv2-128			& dropout			& conv2-128 \\
					& FC-10				& maxpool2d			& FC-200			& batchnorm2d \\
					& Softmax			& ReLU				& ReLU				& ReLU \\ \hline
			4		& 					& dropout			& dropout			& conv2-128 \\
					& 					& FC-200			& FC-100			& ReLU \\
					& 					& ReLU				& ReLU				& maxpool2d \\ \hline
			5		& 					& dropout			& dropout			& dropout \\
					& 					& FC-10				& FC-43				& conv2-256 \\
					& 					& Softmax			& Softmax			& batchnorm2d \\
					& 					& 					& 					& ReLU \\ \hline
			6		&					& 					& 					& conv2-256 \\
					& 					& 					& 					& ReLU \\
					&					& 					& 					& maxpool2d \\ \hline
			7		&					& 					& 					& dropout \\
					&					& 					& 					& FC-1024 \\
					&					& 					& 					& ReLU \\ \hline
			8		&					& 					& 					& FC-512 \\
					&					& 					& 					& ReLU \\ \hline
			9		&					& 					& 					& dropout \\
					&					& 					& 					& FC-10 \\
					&					& 					& 					& Softmax \\\hline
		\end{tabular}
	\end{center}
	\end{table}

\section{Detecting watermarked inputs}
\label{appendix:wm-detection}

We assess if watermarked inputs can be identified such that \adv could remove them from the  $D_\mathcal{A}$ before training the surrogate model.

This defense consists in first training a DNN model with the whole training dataset.
Then, training data is predicted using the trained model and we record the activations of the last hidden layer of the DNN model. These activations are projected to three dimensions using Independent Component Analysis (ICA) and clustered into two clusters using k-means.
These clusters are expected to group benign training data and poisoned data (watermarked inputs) respectively.
The intuition for this approach is that incorrectly labeled inputs (watermark) trigger different activations than correctly labeled inputs in the trained DNN model.
The size and silhouette score~\cite{rousseeuw1987silhouettes} of the two clusters are analyzed to conclude (1) if there is backdoor in the model and (2) which training inputs compose the backdoor. According to authors, a low silhouette score (0.1/0.15) and a high difference in relative cluster size is expected if the model embeds a watermark. The smallest cluster should contain the watermarked inputs.

\setlength\tabcolsep{4pt}
\begin{table}
\begin{center}
	\caption{Results of watermark detection~\cite{chen2018detecting} on several watermarked ($wm$) and plain models (No $wm$). Relative size represents the average ratio of training inputs ($D_\mathcal{A}$) contained in the small cluster (supposed to contain watermarked inputs only). $wm$ split counts watermarked inputs in the small/large clusters ($\vert T_\mathcal{A} \vert = 250$). Silhouette score~\cite{rousseeuw1987silhouettes} is averaged over all classes. Small clusters are much larger than the size of the watermark. Most watermarked inputs are contained in large clusters. Silhouette score for both plain and watermarked models is above the recommended detection threshold (0.15). \ourname's watermark cannot be detected.}
	\label{tab:watermark-detection}
	\resizebox{1.\columnwidth}!{
	\begin{tabular}{lccccc} \hline
						& \multicolumn{2}{c}{Relative size} & $wm$  	& \multicolumn{2}{c}{Silhouette score} \\
		 Model			& $wm$ 		& No $wm$ 				& split 	& $wm$ 			& No $wm$  	\\ \hline
		 MNIST-5L 		& 0.222  	& 0.449					& 55/195 	& $0.47\pm0.33$	& $0.23\pm0.06$ \\
		 GTSRB-5L 		& 0.059  	& 0.098 				& 59/191 	& $0.64\pm0.18$	& $0.76\pm0.20$  \\
		 CIFAR10-9L 	& 0.094  	& 0.105					& 3/247 	& $0.79\pm0.18$	& $0.78\pm0.19$ \\ \hline
		 GTSRB-RN34 	& 0.409  	& 0.419 				& 92/158	& $0.26\pm0.09$ & $0.24\pm0.02$ \\
		 CIFAR10-RN34 	& 0.378  	& 0.338 				& 79/171	& $0.24\pm0.01$ & $0.27\pm0.04$ \\
		 Caltech-RN34 	& 0.425  	& 0.422 				& 124/126	& $0.24\pm0.02$ & $0.24\pm0.02$ \\ \hline
		 Overall 	& 0.264	& 0.305 & 67/183 & 0.44 &  0.42 \\

	\end{tabular}}

\end{center}
\end{table}

\begin{figure}[th]
                \centering
                \includegraphics[width=0.95\columnwidth]{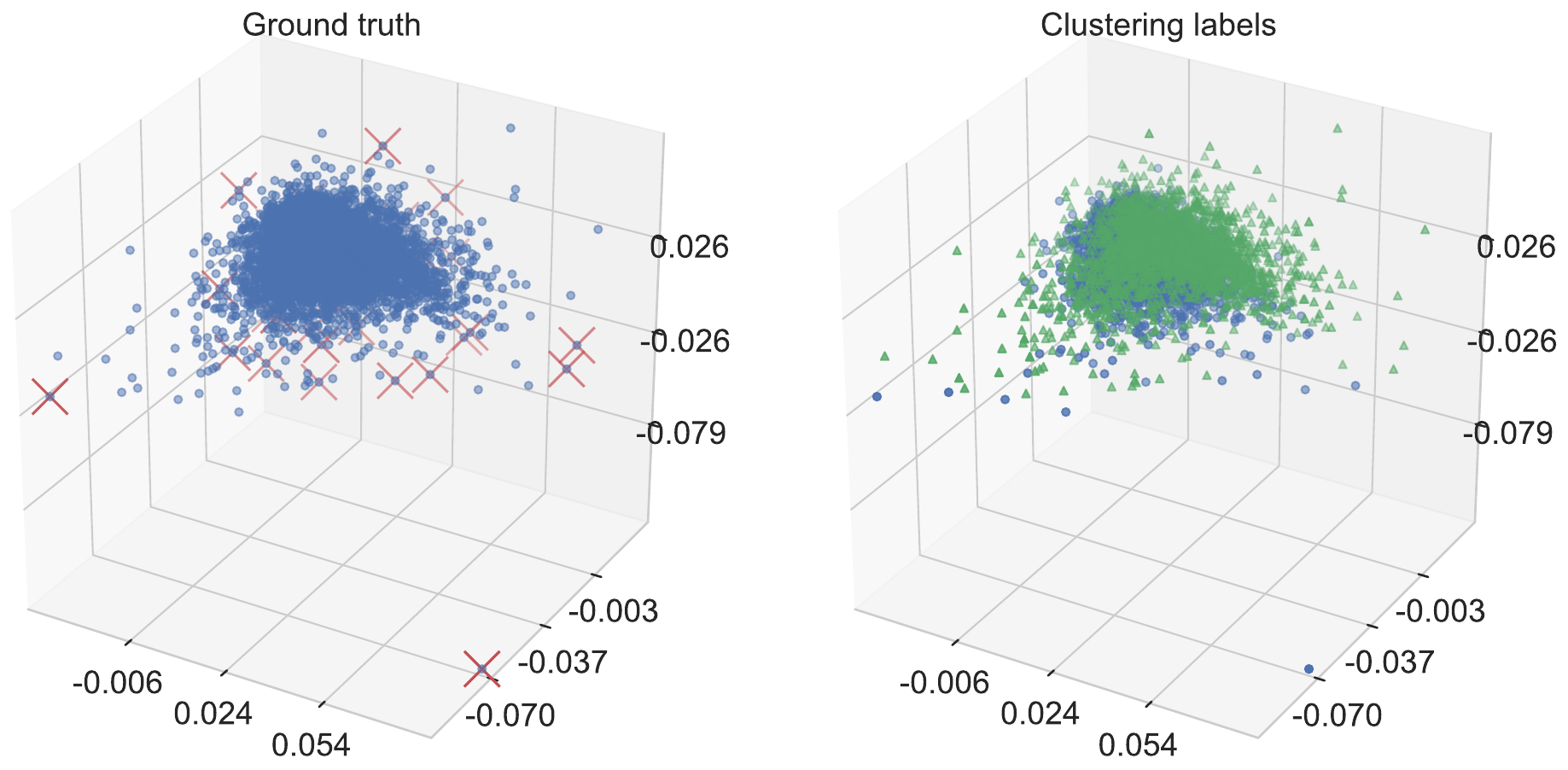}
                \includegraphics[width=0.95\columnwidth]{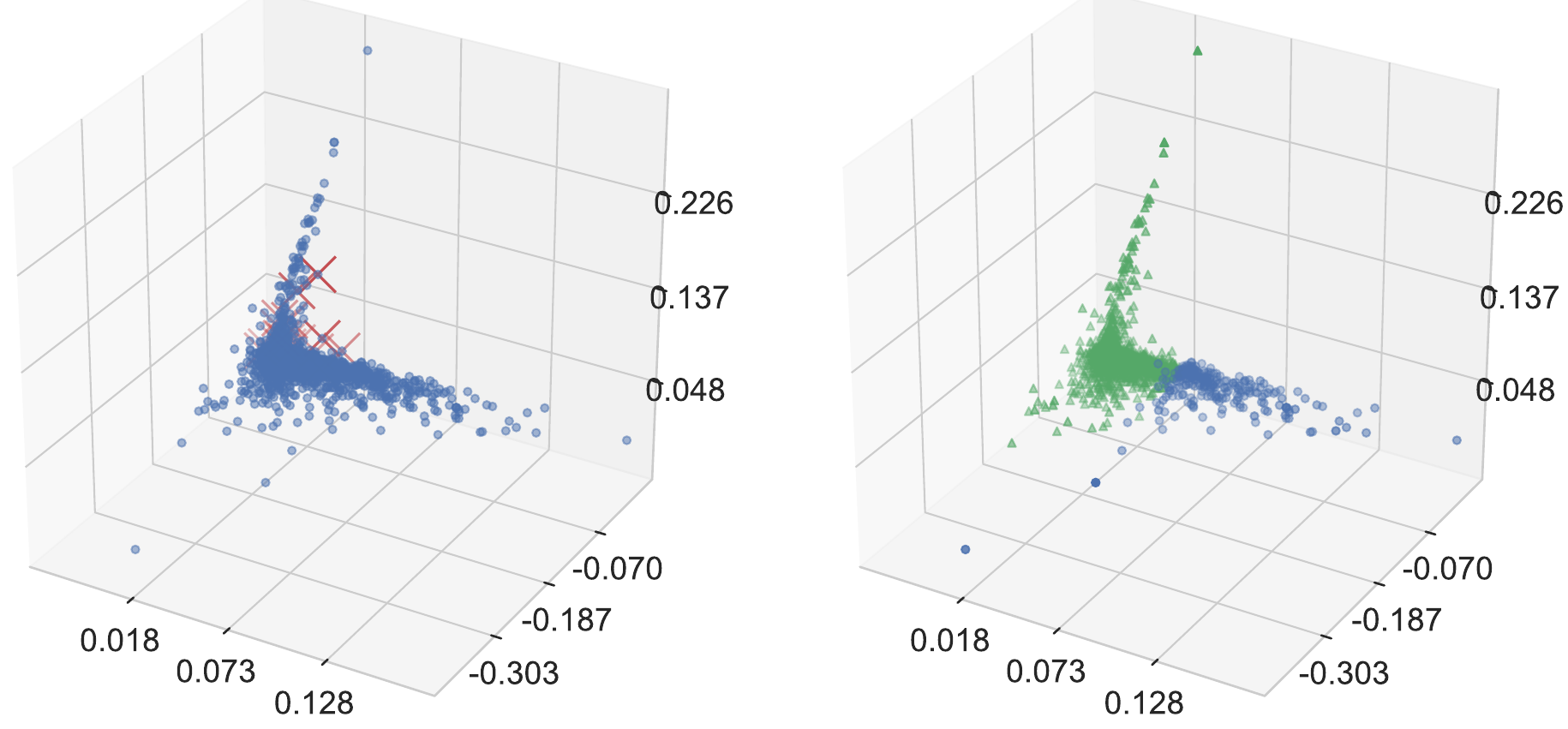}
                \caption{Last activation (three independent components) for inputs predicted ``1'' by MNIST-5L watermarked model (top) and ``4'' by GTSRB-5L watermarked model (bottom). Left: Ground truth watermarked inputs (red cross) and correctly labeled inputs (blue dots). Right: Clustering results of watermark detection (two clusters: blue dots / green triangles). Watermarked inputs are mixed with correctly labeled inputs and clusters cannot isolate the watermark.}
                \label{fig:watermark-detection}
\end{figure}

To evaluate this defense against \ourname, we trained two sets of DNN models, plain models using a correctly labeled training set only, and watermarked models each embedding a watermark of size $\vert T_\mathcal{A} \vert = 250$.
We applied the watermark detection process discussed above on these models and report results in Tab.~\ref{tab:watermark-detection}.
Clustering is not able to isolate watermarked inputs into a single cluster; the main part of watermarked inputs belongs to large clusters.
The recommendation~\cite{chen2018detecting} to discard small clusters from training would deprive $D_\mathcal{A}$ from a large number of correctly labeled samples while a large part of the watermark would be preserved.
Using this approach, the defense wrongly discards 26.4\% of clean data from $D_\mathcal{A}$ on average while detecting only 67 out of 250 watermarked samples (26.8\% of $T_\mathcal{A}$).
The silhouette score is not useful for detecting the watermark either since watermarked and plain models have close scores that are all above the recommended detection threshold (0.1/0.15)~\cite{chen2018detecting}. Our plain models are detected as embedding a watermark using this defense.
We conclude that this defense is ineffective at detecting watermarks generated by \ourname.

We assume the reason for this ineffectiveness is due to the nature of our watermark, selected from the same distribution as the training set.
In contrast to prior DNN watermarking solutions~\cite{merrer2017adversarial,adi2018turning,zhang2018protecting}, our watermarked inputs do not come from a single manifold distant from the training data manifold.
Consequently the model does not learn a ``single'' activation that generalizes to the whole watermark but rather learns individual exceptions for each watermarked input.
The activations of watermarked inputs are thus different from each other and they are scattered among the activations of the remaining training data (correctly labeled).
This can be observed in Fig.~\ref{fig:watermark-detection} - left, where we see that watermarked inputs are scattered among correctly labeled inputs. This explains why generated clusters cannot isolate watermarked inputs from correctly labeled inputs (Fig.~\ref{fig:watermark-detection} - right).

\end{document}